\documentclass[preprint,12pt,3p]{elsarticle}
\usepackage{geometry}                
\usepackage{graphicx}
\usepackage{amssymb, amsmath}
\usepackage{epstopdf}
\usepackage{subfigure}

\usepackage{url}

\graphicspath{{figsupdated/}}

\begin{document}

\begin{frontmatter}

%
%
%
%

\title{Computing via material topology optimisation}

\author[AS]{Alexander Safonov}
\address[AS]{Centre for Design, Manufacturing and Materials, Skolkovo Institute of Science and Technology, Moscow, Russia}
\ead{a.safonov@skoltech.ru}

\author[AA]{Andrew Adamatzky}
\address[AA]{Unconventional Computer Centre, University of the West of England, Bristol, UK}
\ead{andrew.adamatzky@uwe.ac.uk}

\begin{abstract}
We construct logical gates via topology optimisation (aimed to solve a station problem of heat conduction) 
of a conductive material layout. Values of logical variables are represented high  and low  values of a temperature at given sites. Logical functions are implemented via the formation of an optimum layout of conductive material between the sites 
with loading conditions. We implement  {\sc and} and {\sc xor} gates and a one-bit binary half-adder. \\
\end{abstract}


\end{frontmatter}

\section{Introduction}

Any programmable response of  a material to external stimulation can be interpreted as computation. 
To implement a logical function in a material one must map space-time dynamics of an internal structure of a material onto  a space of logical values. 
This is how experimental laboratory prototypes of unconventional computing devices are made: logical gates, circuits and binary adders employing interaction of wave-fragments in light-sensitive Belousov-Zhabotinsky media \cite{ref1}, swarms of soldier crabs \cite{ref2}, growing lamellipodia of slime mould  Physarum polycephalum~\cite{adamatzky2016logical}, crystallisation patterns in ``hot ice'' \cite{adamatzky2009hot},  peristaltic waves in protoplasmic tubes \cite{adamatzky2014slime}.  In many cases logical circuits are `built'  or evolved from a previously disordered material \cite{miller2014evolution}, e.g. networks of slime mould \emph{Physarum polycephalum} \cite{whiting2016towards}, bulks of nanotubes \cite{broersma2012nascence}, nano particle ensembles \cite{bose2015evolution, broersma2017computational}. In these works the computing structures could be seen as growing on demand, and logical gates develop in a continuum where an optimal distribution of material minimised internal energy.  A continuum exhibiting such properties can be coined as a ``self-optimising continuum''. Slime mould of \emph{Physarum polycephalum} well exemplifies such a continuum: the slime mould is capable of solving many computational problems, including mazes and adaptive networks  \cite{adamatzky2016advances}. Other examples of the material behaviour include bone remodelling \cite{christen2014bone}, 
roots elongation \cite{mazzolai2010plant}, sandstone erosion \cite{bruthans2014sandstone}, 
crack and lightning propagation \cite{achtziger2000optimization}, growth of neurons and blood vessels etc. 
Some other physical systems suitable for computations were also proposed in \cite{miller2014evolution, turner2014neuroevolution, banzhaf2006guidelines,  miller2002evolution}. In all these cases, a phenomenon of the formation of an optimum layout of material is related to non-linear laws of material behaviour, 
resulting in the evolution of material structure governed by algorithms similar to those used in a topology optimisation of structures~\cite{klarbring2010dynamical}. We develop the ideas of material optimisation further and show, in numerical models, how logical circuits can be build in a conductive material self-optimise its structure governed by configuration of inputs and outputs. 

The paper is structured as follows. In Sect.~\ref{topologyoptimisation} we introduce  topology optimisation aimed to solve a problem of a stationary heat conduction.
Gates {\sc and} and {\sc xor} are designed and simulated in Sects.~\ref{andgate} and \ref{xorgate}. We design one-bit half-adder in Sect.~\ref{onebithalfadder}.
Directions of further research are outlined in Sect.~\ref{discussion}.

\section{Topology optimisation}
\label{topologyoptimisation}

A topology optimisation in continuum mechanics aims to find a layout of a material within a given 
design space that meets specific optimum performance targets \cite{bendsoe2013topology, hassani2012homogenization, huang2010evolutionary}.  
The topology optimisation is applied to solve a wide range of problems \cite{bendsoe2005topology}, 
e.g. 
maximisation of heat removal for a given amount of  heat conducting material \cite{bejan1997constructal}, 
maximisation of fluid flow within channels \cite{borrvall2003topology},
maximisation of structure stiffness and strength \cite{bendsoe2005topology}, 
development of meta-materials satisfying specified  mechanical and thermal physical properties \cite{bendsoe2005topology}, 
optimum layout of plies in composite laminates \cite{stegmann2005discrete}, 
the design of an inverse acoustic horn \cite{bendsoe2005topology}, modelling of amoeboid organism growing towards food sources \cite{Safonov20161},
optimisation of photonics-crystal band-gap structures \cite{men2014robust}.

A standard method of the topology optimisation employs a modelling material layout that uses a density of material, $\rho$, 
varying from 0 (absence of a material) to 1 (presence of a material),  where a dependence of structural properties on the density 
of material is described by a power law. This method is known as Solid Isotropic Material with Penalisation (SIMP) \cite{zhou1991coc}. 
An optimisation of the objective function consists in finding an optimum distribution of $\rho$:  $\min_\rho f(\rho)$.

The problem can be solved in various numerical schemes, including the sequential quadratic programming (SQP) \cite{wilson1963simplicial}, 
the method of moving asymptotes (MMA) \cite{svanberg1987method}, and the optimality criterion (OC) method \cite{bendsoe2005topology}. 
The topology optimisation problem can be replaced with a problem of finding a stationary point of an Ordinary Differential Equation (ODE) \cite{klarbring2010dynamical}. Considering density constraints
on $\rho$, the right term of ODE is equal to a projection of the negative gradient of the objective function. Such
optimisation approach is widely used in the theory of projected dynamical systems \cite{nagurney2012projected}. Numerical schemes of topology
optimisation solution can be found using simple explicit Euler algorithm. As shown in \cite{klarbring2012dynamical} iterative schemes match the
algorithms used in bone remodelling literature \cite{harrigan1994bone}.

In this work the topology optimisation problem as applied to heat conduction problems \cite{gersborg2006topology}. 
Consider a region in the space $\Omega$ with a boundary  
$\Gamma=\Gamma _D \cup \Gamma _N$,  $\Gamma _D \cap \Gamma_N= \emptyset$,  
separated for setting the Dirichlet (D) and the Neumann (N) boundary conditions. 
For the region $\Omega $  we consider the steady-state heat equation given in: 

\begin{equation}
\nabla \cdot k \nabla T +f=0  \text{ in } \Omega 
\end{equation}

\begin{equation}
T = T_0  \text{ on } \Gamma_D 
\end{equation}

\begin{equation}
(k \nabla T) \cdot n = Q_0 \text{ on } \Gamma_N 
\end{equation}

where  $T$ is a temperature, 
$k$ is a heat conduction coefficient, 
$f$ is a volumetric heat source, and 
$n$ is an outward unit normal vector. 
At the boundary $\Gamma_D$  a temperature  $T=T_0$ is specified in the form of Dirichlet
boundary conditions, and at the boundary  $\Gamma _N$ of  the heat flux $(k \nabla T) \cdot n$ is specified using Neumann boundary conditions. 
The condition $(k \nabla T) \cdot n = 0$   specified at the part of  $\Gamma_N$  means a thermal insulation  (adiabatic conditions).

When stating topology optimisation problem for a solution of the heat conduction problems it is necessary to find an optimal distribution for a limited volume of conductive material in order to minimise heat release, which corresponds to designing a thermal conductive device. It is necessary to find an optimum distribution of material density $\rho$  within a given area  $\Omega$ in order to minimise the cost function:

\begin{equation}
\text{Minimize } C(\rho) = \int_\Omega \nabla T \cdot (k (\rho) \nabla T)
\end{equation}

\begin{equation}
\text{Subject to } \int_\Omega \rho <M 
\end{equation}

In accordance with the SIMP method the region being studied can be divided into finite elements
with varying material density $\rho_i$ assigned to each finite element $i$. A relationship
between the heat conduction coefficient and the density of material is described by a power law as follows:

\begin{equation}
k_i = k_{\min} + (k_{\max} - k_{\min}) \rho^p_i ,  \hspace{5mm}  \rho_i \in \lfloor 0, 1 \rfloor
\end{equation}
where  
$k_i$ is  a value of heat conduction coefficient at the $i$-th finite element,
$\rho_i$ is a  density value at the $i$-th element,
$k_{\max}$ is a heat conduction coefficient at $\rho_i=1$, 
$k_{\min}$ is a heat conduction coefficient at $\rho_i=0$,
$p$ is a penalisation power ($p>1$).

In order to solve the problem (1)--(6) we apply the following techniques used in the dynamic systems modelling. 
Assume that  $\rho$  depends on a time-like variable $t$.  
Let us consider the following differential equation to determine density in $i$-th finite element,  $\rho_i$, 
when solving the problem stated in (1)--(6):

\begin{equation}
\acute{\rho_i}=\lambda (\frac{C_i(\rho_i)}{\rho_i V_i} - \mu), \hspace{5mm}  C_i(\rho_i) = \int_{\Omega_i} \nabla T \cdot (k_i(\rho) \nabla T) d\Omega
\end{equation}
where dot above denotes the derivative with respect to $t$, 
$\Omega_i$ is a  domain of  $i$-th finite element, 
$V_i$ is a volume of $i$-th element, 
$\lambda$ and $\mu$ are positive constants characterising behaviour of the model. 
This equation can be obtained by applying methods of the projected dynamical systems \cite{klarbring2012dynamical} 
 or bone remodelling methods \cite{harrigan1994bone, mullender1994physiological, payten1998optimal}.

For numerical solution of equation (8) a projected Euler method is used \cite{nagurney2012projected}. 
This gives an iterative formulation for  the solution finding $\rho_i$ \cite{klarbring2010dynamical}:
\begin{equation}
\rho^{n+1}_i = \rho^n_i + q[\frac{C_i(\rho^n_i)}{\rho^n_i V_i} - \mu^n]
\end{equation}
where $q = \lambda \Delta t$, 
$\rho^{n+1}_i$ and $\rho^n_i$ are the numerical approximations of  $\rho_i(t+\Delta t)$ and $\rho_i(t)$,   
$\mu^n =\frac{\sum_i C_i(\rho^n_i)}{\sum_i \int_{\Omega_i} \rho_{ev} d\Omega}$, $\rho_{ev}$ is a specified mean value of density.

We consider a modification of equation (8):
\begin{equation}
\rho^{n+1}_i = 
\begin{cases}
\rho^n_i + \theta \text{ if } \frac{C_i(\rho^n_i)}{\rho^n_i V_i} - \mu^n \geq 0, \\
 \rho^n_i - \theta \text{ if } \frac{C_i(\rho^n_i)}{\rho^n_i V_i} - \mu^n < 0,
\end{cases}
\end{equation}
where $\theta$ is a positive constant.

Then we calculate a value of $\rho _i^{n+1}$  using equation (9) and project $\rho _i$ onto a set of constraints:

\begin{equation}
\rho^{n+1}_i = 
\begin{cases}
\rho_{\max} \text{ if } \rho^{n+1}_i > \rho_{\max}, \\
\rho^{n+1}_1 \text{ if } \rho_{\min} \leq \rho^{n+1}_i \leq \rho_{\max},\\
\rho_{\min} \text{ if } \rho^{n+1}_i < \rho_{\min} 
\end{cases}
\end{equation}
where  $\rho_{\min}$ is a specified minimum value of $\rho_i$ and $\rho_{\max}$ is a specified maximum value of $\rho_i$. 
A minimum value is taken as the initial value of density for all finite elements:  $\rho_i^0=\rho_{\min}$.

\section{Specific parameters}
\label{Specificparameters}

The algorithm above is implemented in ABAQUS \cite{Abaqus2014} using the modification of the structural topology optimisation plug-in, 
UOPTI, developed previously \cite{Safonov2015}. Calculations were performed using topology optimisation methods for the finite element model
of $200 \times 200 \times 1$ elements. Cube-shaped linear hexahedral elements of DC3D8 type with a unit
length edges were used in calculations. The elements used have eight integration points. The cost function value is
updated for each finite element as a mean value of integration points for an element under consideration  \cite{Abaqus2014}.

The model can be described by the following parameters: 
$\rho_{\min}$ and  $\rho_{\max}$ are minimum and maximum values of $\rho_i$, 
$M=\sum_i \int_{\Omega_i} \rho_{ev} d\Omega$ is a mass of the conductive material, 
$\theta$ is an increment of $\rho_i$ at each time step, 
$p$ is a penalisation power, 
$k_{\max}$ is a heat conduction coefficient at $\rho_i = 1$, 
$k_{\min}$ is a heat conduction coefficient at $\rho_i = 0$. 
All parameters but $M$ are the same for all six (three devices with two type of boundary conditions) implementations:
$\rho_{\max}=1$,
$\rho_{\min}=0.01$,
$\theta=0.03$,
$p=2$,
$K_{\max}=1$,
$K_{\min}=0.009$. 

The parameter $M$ is specified as follows: 
$M=2000$ for {\sc and}, {\sc xor} in Dirichlet boundary conditions on inputs, and one-bit half-adder for both types of boundary conditions;
$M=800$ for {\sc and} gate and $M=400$ for {\sc xor} gate in Neumann boundary conditions. 

We use the following notations. Input logical variables are $x$ and $y$, output logical variable is $z$. 
They takes values 0 ({\sc False}) and 1 ({\sc True}).
Sites in input stimuli the simulated material are $I_x$ and $I_y$ (inputs),  $O$, $O_1$, $O_2$ (outputs). 
Sites of outlets are $V$, $V_1$ and $V_2$ (temperature are set to 0 in the outlet, 
so we use symbol $V$  by analogy with vents in fluidic devices).  Temperature at the sites is shown as 
$T_{I_x}$, $T_{I_y}$, $T_{O}$, $T_{O_1}$ etc. We show distances between as $l(I_x, I_y)$, $l(I_x, O)$ etc.

Logical values are represented by temperature: $x=1$ is $T_{I_x}=100$ and $x=0$ is  $T_{I_x}=0$, the same for $y$. 
We input data in the gates by setting up thermal boundary conditions are set at the input sites and  adiabatic boundary conditions 
for other nodes. The temperature at each point is specified by setting equal values in 4 neighbour nodes belonging to the same finite 
element. Temperature at outputs and outlets is set to zero of all experiments: $T_O=T_{O_1}=T_{O_2}=0$,
 $T_V=T_{V_1}=T_{V_2}=0$. To maintain specified boundary conditions we setup a thermal flow  through the boundary points. Intensity of the flows is determined via solution of the thermal conductivity equation at each iteration. Therefore  intensity of the thermal streams via input, output and outlet sites changes during the simulation  depending on a density distribution of the conductive material. Namely, if we define zero temperature at a site the intensity of the stream though the site will be negative if a density of the conductive material is maximal; the intensity will be zero if the material density is minimal. In case when we do not define a temperature at a site the intensity is non-zero if the density is maximal and zero if the density is minimal.
 Therefore, instead of talking about temperature at the output we talk about thickness of the conductive material. 
 Namely, if the material  density value at the output site $O$ is minimal, $\rho_O = \rho_{\min}$, we assume logical output 0 ({\sc False}). 
 If the density $\rho_O = \rho_{\max}$ we assume logical output 1 ({\sc True}).  The material density for all finite elements is set to a
minimum value $\rho _i^0=\rho _{\min}$ at the beginning of computation. 

 In case of Dirichlet boundary conditions in inputs, in $x=0$ and $y=0$  the temperature is constant and equal to zero at all points, therefore the temperature gradient is also zero, $\nabla T=0$. The cost function is also equals to zero at all points: $C_i(\rho _i)=0$. 
As the initial density for all finite elements is set to a minimum value $\rho_i^0=\rho_{\min}$ then 
 from equations (9) and (10) follows that the density stays constant and equal to its minimum value  
 $\rho_i^n=\rho_{\min}$.  Therefore, the density value at  $O$ point is minimal,  $\rho_O=\rho_{\min}$ which indicates 
 logical output 0. Further we consider only situations when one of the inputs is non-zero.
 
 In case of Neumann boundary conditions in inputs a flux in each site is specified by setting the flux through the face of the finite element to which the site under consideration belongs.  Adiabatic boundary conditions are set for other nodes. The logical value of $x$ is represented by the value of given flux in $I_x$, $Q_{I_x}$. The logical value of $y$ is represented by the value of given flux in $I_y$, $Q_{I_y}$. Flux $Q_{I_x}=0$ represents $x=0$ and flux $Q_{I_x}=1$ represents $x=1$. 
 
 Figures in the paper show density distribution of the conductive material. The maximum values of $\rho$ are shown by red colour, 
 the minimum values by blue colour.

\section{{\sc and} gate}
\label{andgate}

\subsection{Dirichlet boundary conditions}

\begin{figure}[!tbp]
\centering
\subfigure[]{\includegraphics[scale=0.25]{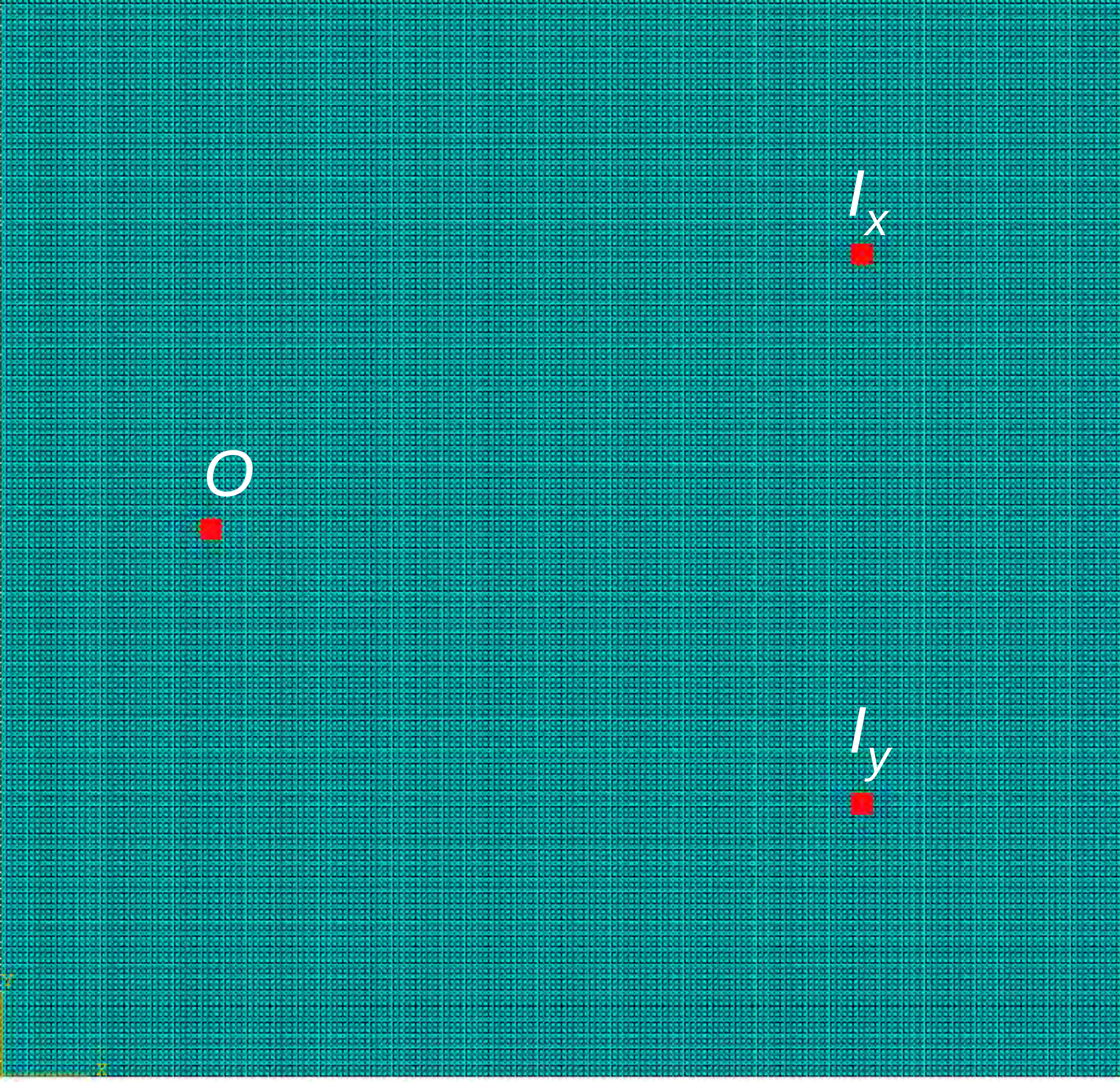} }
\subfigure[]{\includegraphics[scale=0.25]{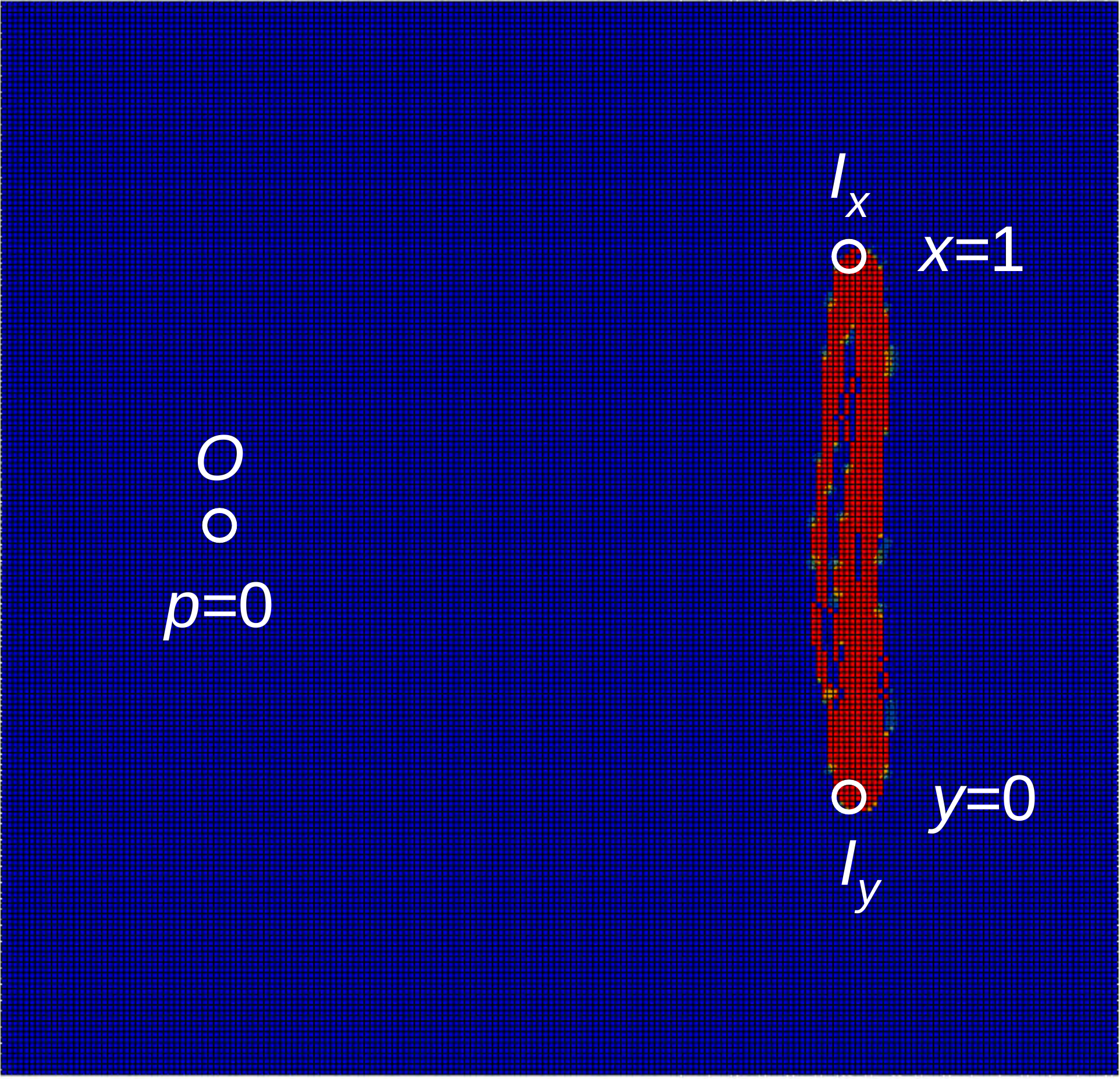}}
\subfigure[]{\includegraphics[scale=0.25]{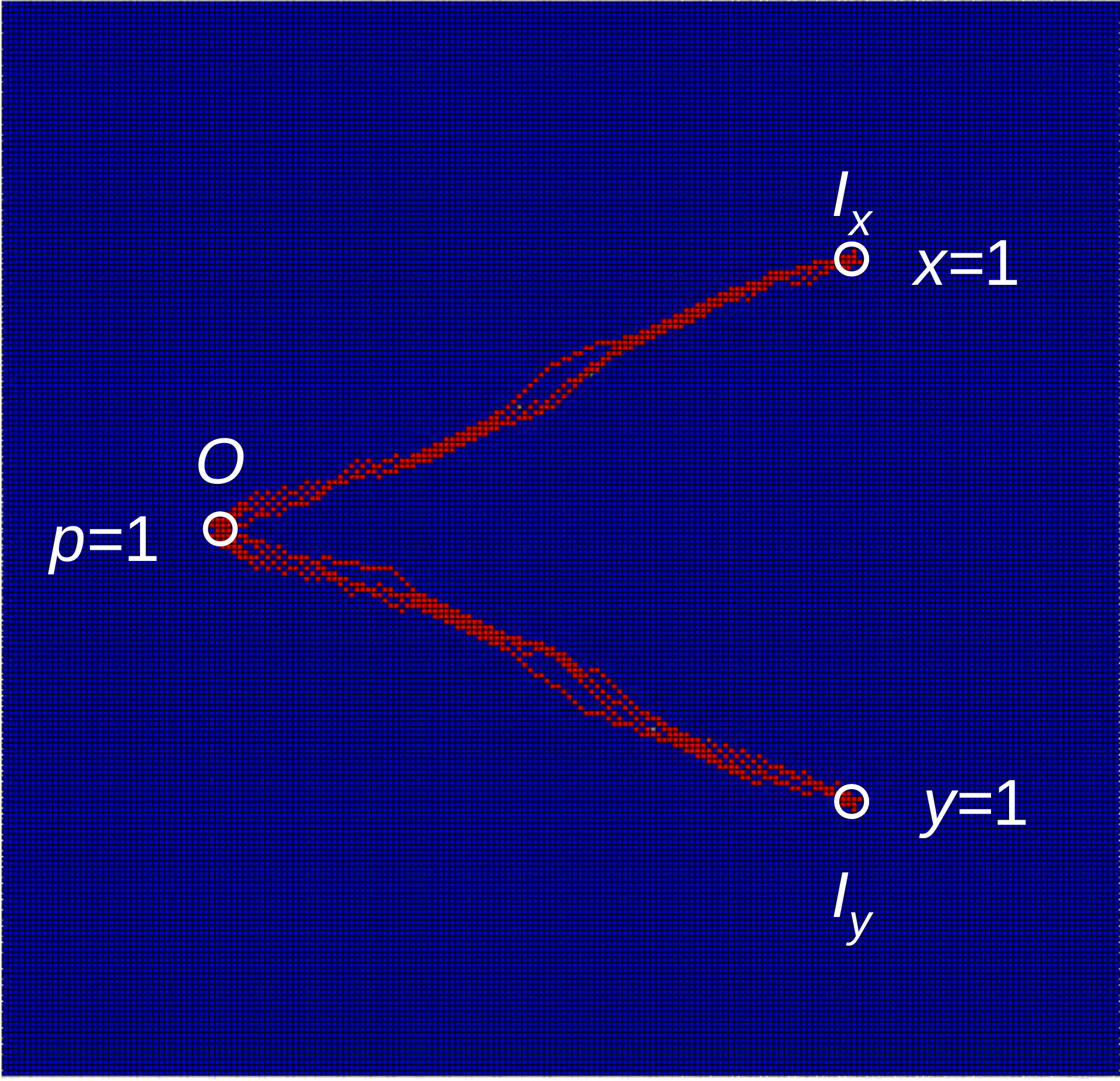}}
\caption{{\sc and} gate implementation with Dirichlet boundary conditions. 
(a) Scheme of inputs and outputs. (bc) Density distribution $\rho$ for inputs (b) $x=1$ and $y=0$ and 
(c ) $x=1$ and $y=1$. }
\label{fig1}
\end{figure}

\begin{figure}[!tbp]
\centering
\subfigure[$t=10$]{ \includegraphics[scale=0.85]{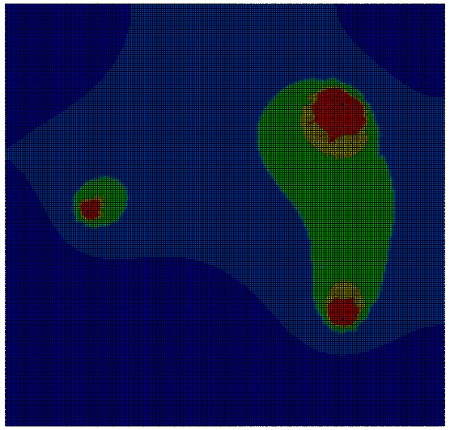} }
\subfigure[$t=20$]{ \includegraphics[scale=0.85]{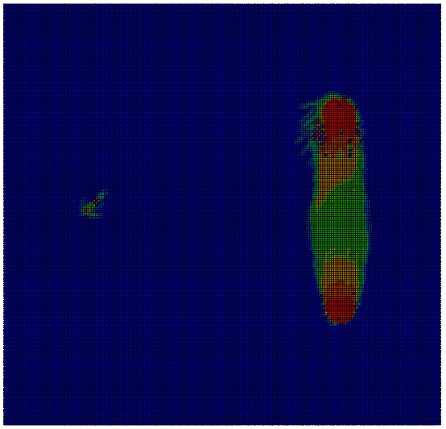} }
\subfigure[$t=30$]{ \includegraphics[scale=0.85]{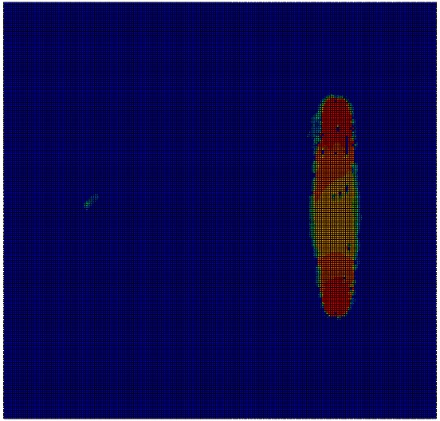} }
\subfigure[$t=40$]{ \includegraphics[scale=0.85]{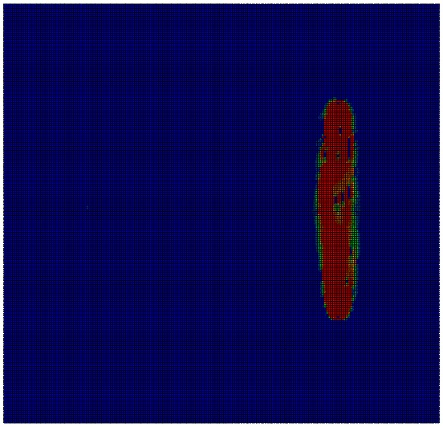} }
\subfigure[$t=50$]{ \includegraphics[scale=0.85]{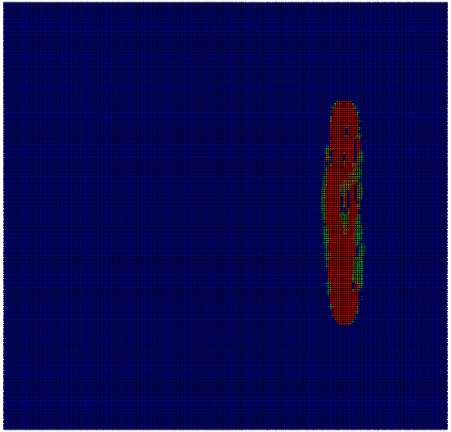} }
\subfigure[$t=100$]{ \includegraphics[scale=0.85]{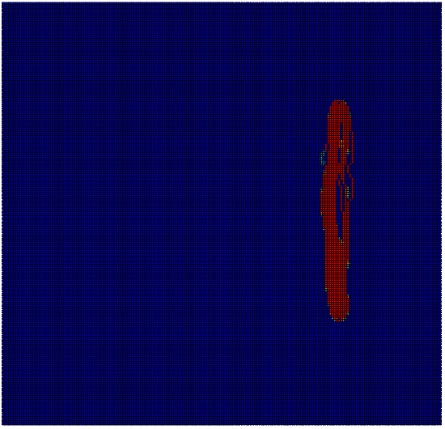} } 
\caption{Density distribution,  $\rho$, in the implementation of {\sc and} gate for inputs  $x=1$  and  $y=0$, 
Dirichlet boundary conditions for input points. The snapshots are taken at t=10, 20, 30, 40, 50, and 100 steps.}
\label{fig2}
\end{figure}

\begin{figure}[!tbp]
\centering
\subfigure[$t=10$]{ \includegraphics[scale=0.85]{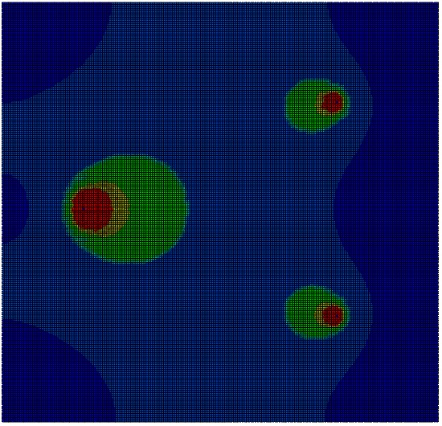} }
\subfigure[$t=20$]{ \includegraphics[scale=0.85]{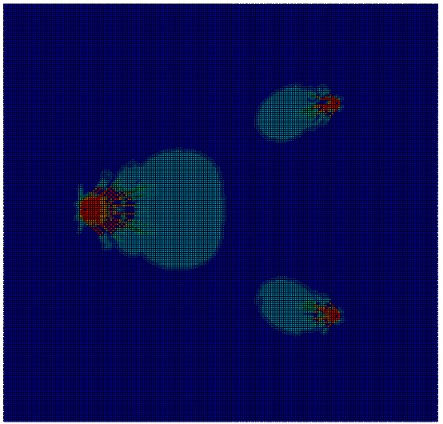} }
\subfigure[$t=30$]{ \includegraphics[scale=0.85]{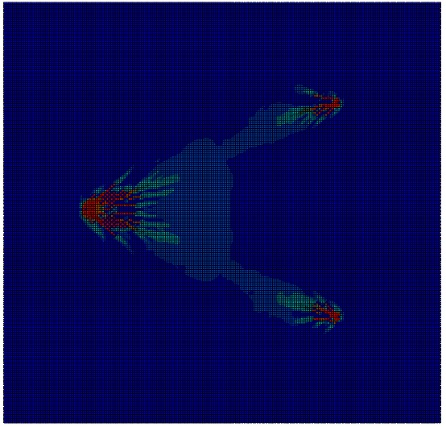} }
\subfigure[$t=40$]{ \includegraphics[scale=0.85]{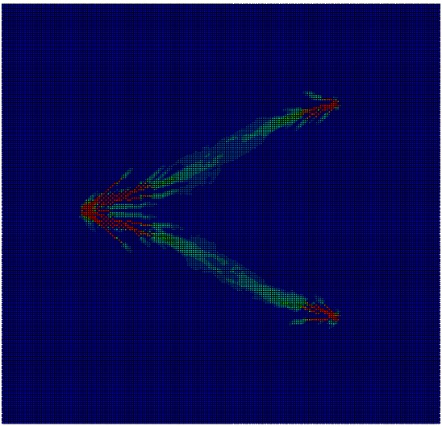} }
\subfigure[$t=50$]{ \includegraphics[scale=0.85]{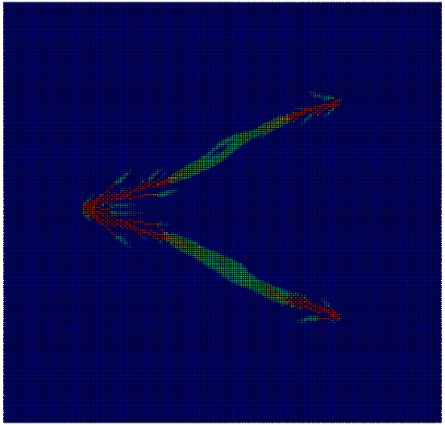} }
\subfigure[$t=100$]{ \includegraphics[scale=0.85]{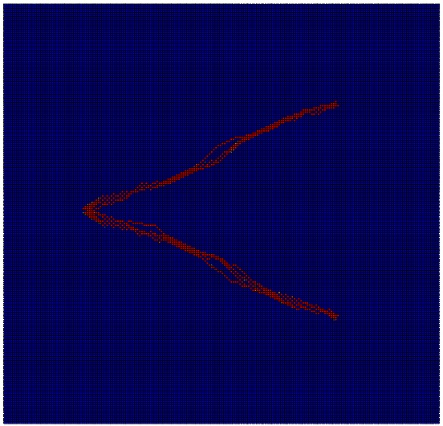} }  
\caption{Density distribution,  $\rho$, in the implementation of {\sc and} gate for inputs  $x=1$  and  $y=1$, 
Dirichlet boundary conditions for input points. The snapshots are taken at t=10, 20, 30, 40, 50, and 100 steps.}
\label{fig3}
\end{figure}

Let us consider implementation of a {\sc and} gate in case of the Dirichlet boundary conditions in the input sites. 
The input $I_x$ and $I_y$ and output $O$ sites are arranged at the vertices of an isosceles triangle (Fig.~\ref{fig1}a):
$l(I_x, I_y)=102$, $l(I_x, O)=127$, $l(I_y, O)=127$. The Dirichlet boundary conditions are set to $I_x$, $I_y$ and $O$.
The material density distribution for inputs $x=1$ and $y=0$ is shown in Fig.~\ref{fig1}b. The maximum density region connects 
$I_x$ with $I_y$ and no material is formed at site $O$, thus output is 0. The space-time dynamics of the gate is shown in Fig.~\ref{fig2}. 
When both inputs are {\sc True}, $x=1$ and $y=1$, domains with maximum density of the material span input sites with output site, 
($I_x, O$) and ($I_y, O$) (Fig.~\ref{fig1}c). Therefore the density value at the output is maximal, $\rho_O = \rho_{\max}$ which indicated 
logical output 1 ({\sc True}). Figure~\ref{fig3} shows intermediate results of density distribution in the gate for $x=1$ and $y=1$. Supplementary videos can be found here \cite{Safonov2016}.

\subsection{Neumann boundary conditions.}

\begin{figure}[!tbp]
\centering
\subfigure[]{\includegraphics[scale=0.25]{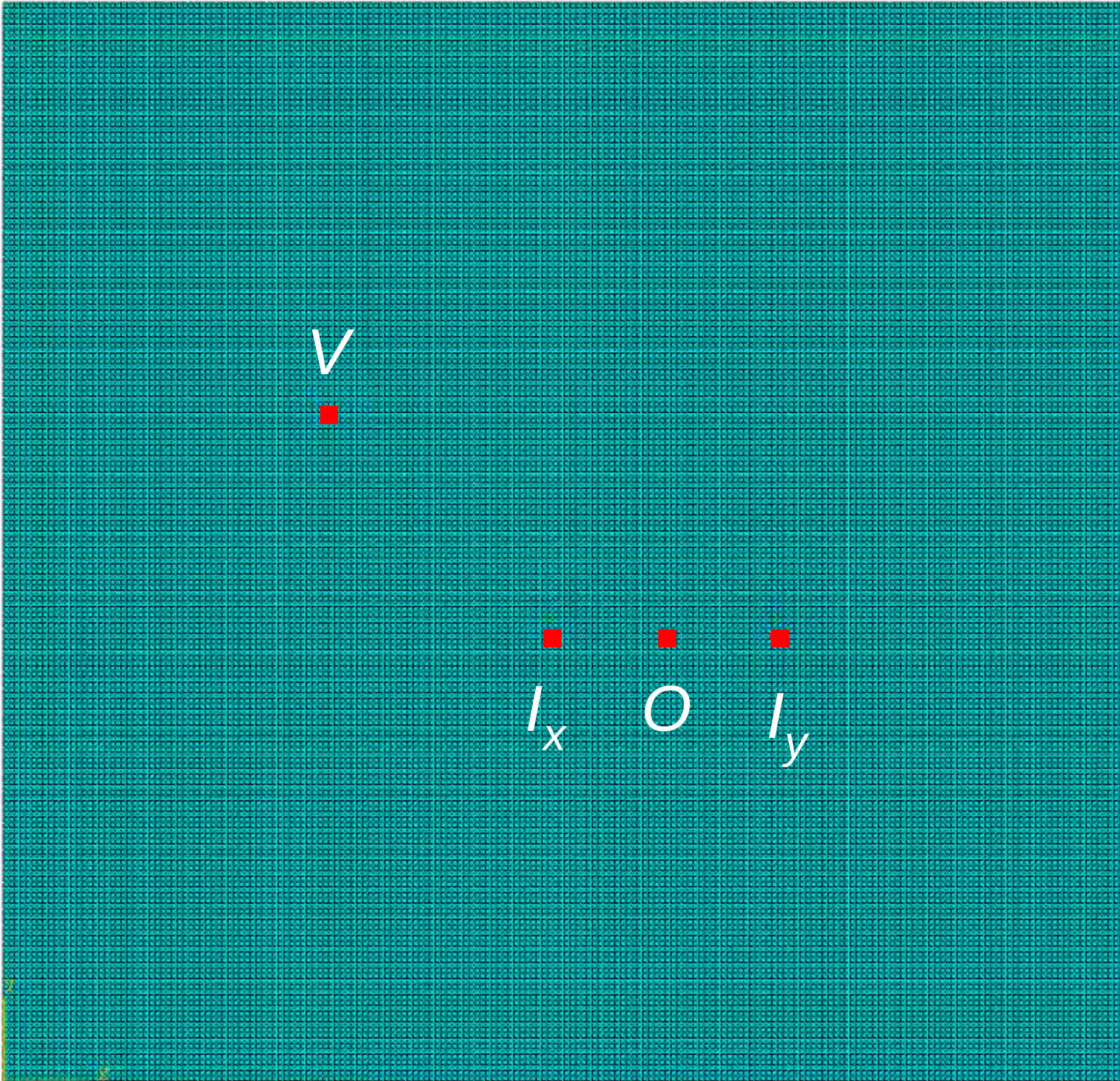}}
\subfigure[]{\includegraphics[scale=0.25]{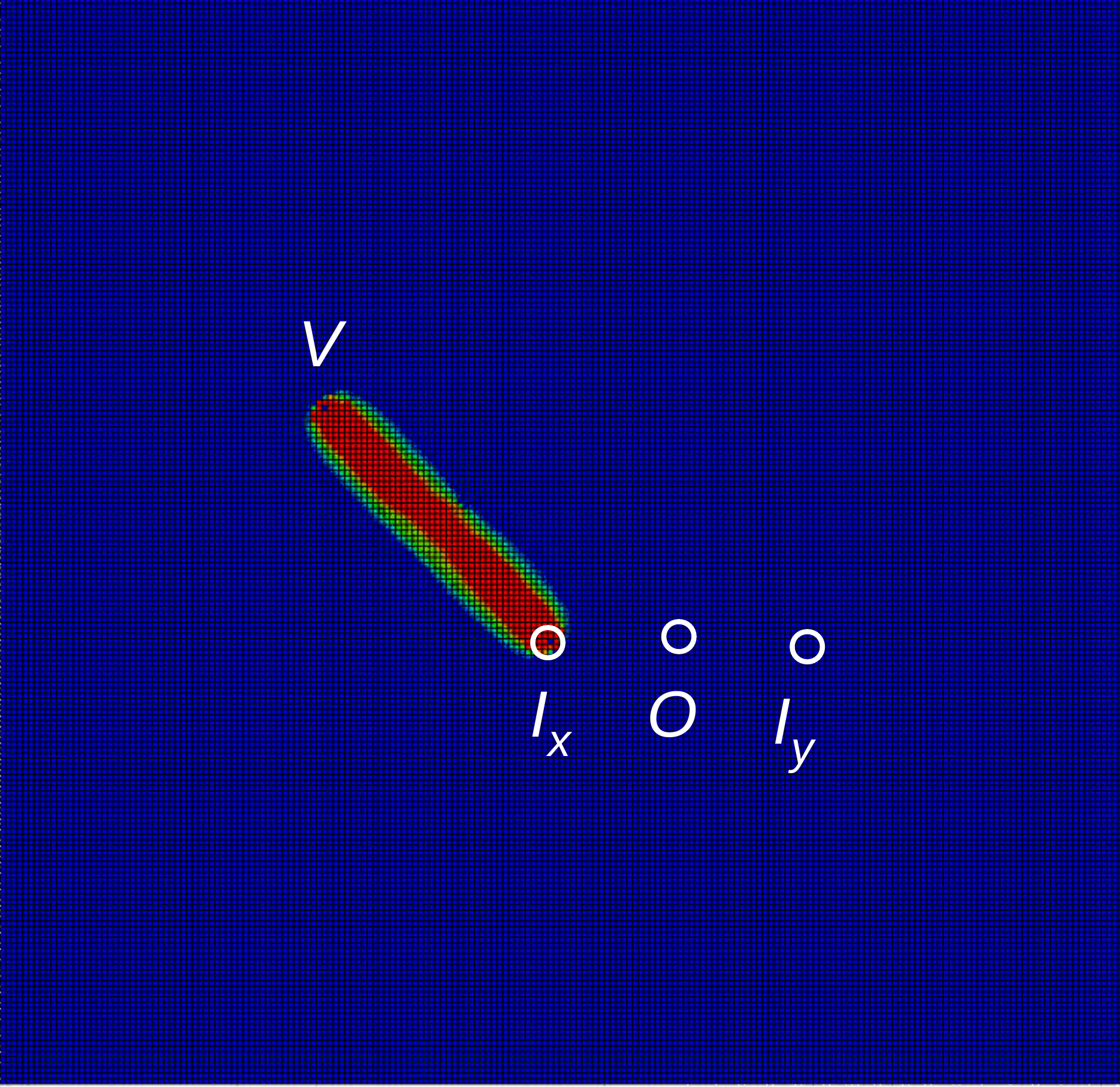}}\\
\subfigure[]{\includegraphics[scale=0.25]{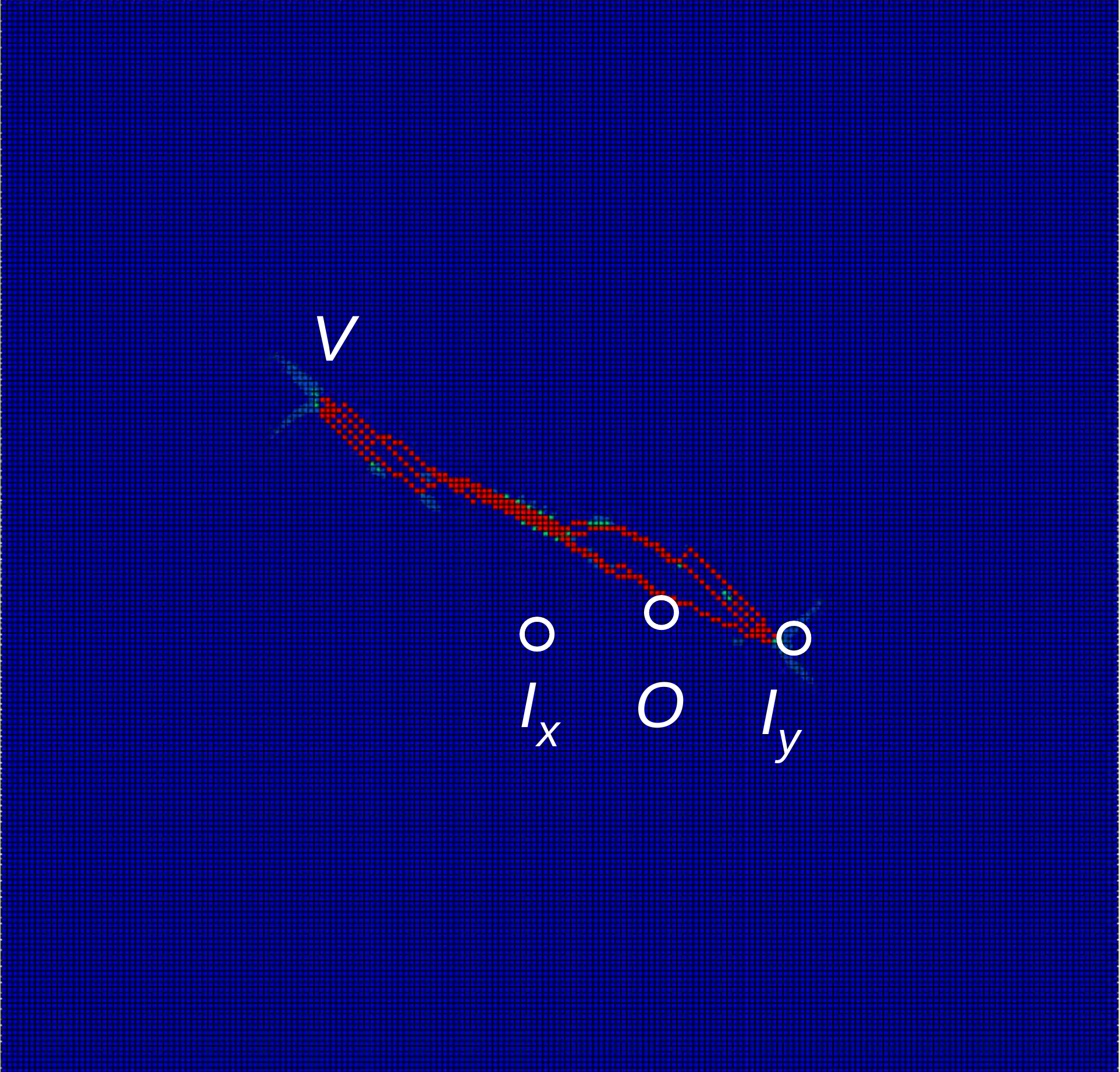}}
\subfigure[]{\includegraphics[scale=0.25]{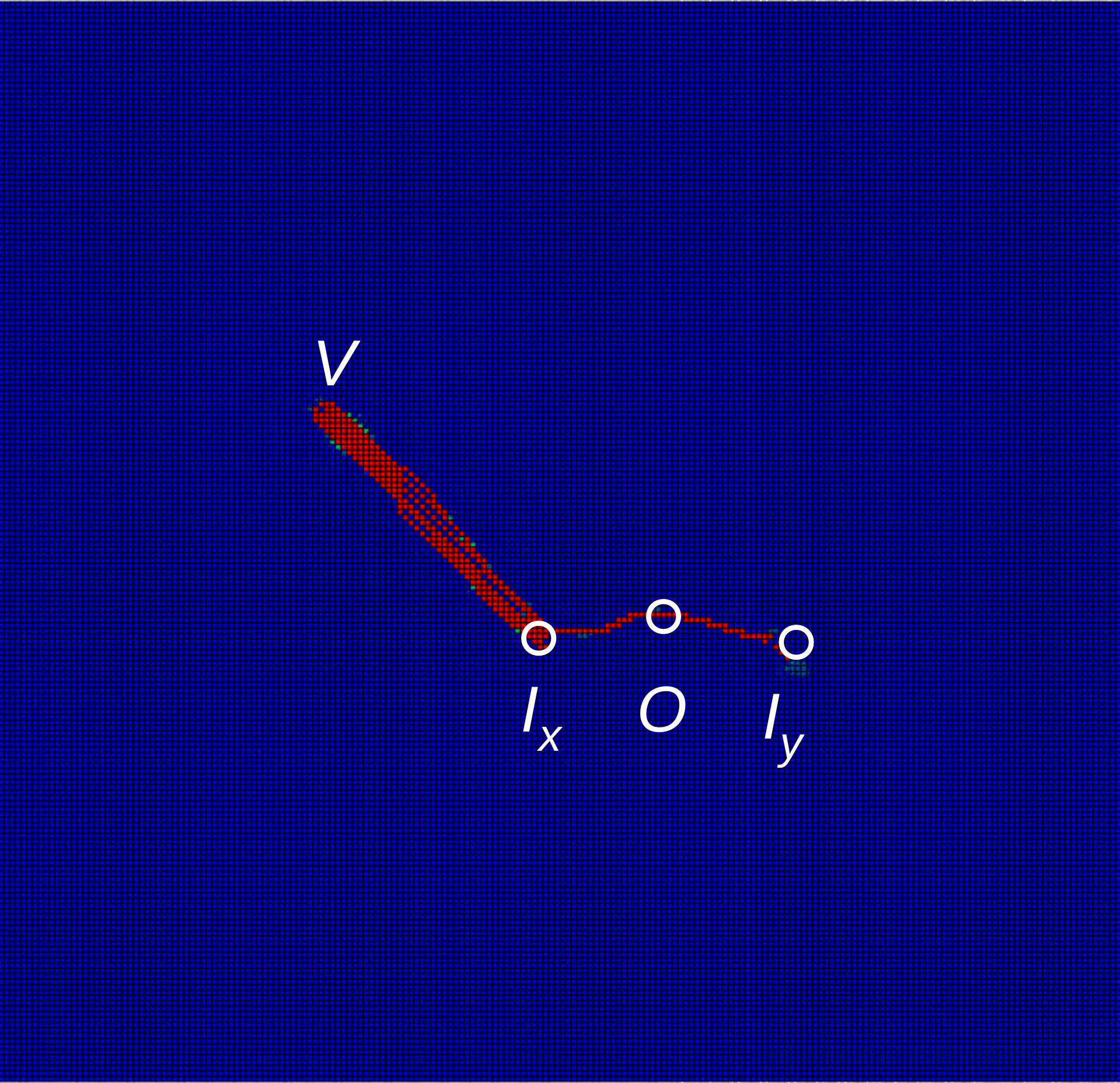}}
\caption{{\sc and} gate implementation in case of Neumann boundary conditions.
(a) Scheme of the gate. Density distribution, $\rho$, for inputs
(b)  $x=1$, $y=0$,
(c ) $x=0$, $y=1$,
(d)  $x=1$, $y=1$.}
\label{fig5}
\end{figure}

\begin{figure}[!tbp]
\centering
\subfigure[$t=10$]{ \includegraphics[scale=0.85]{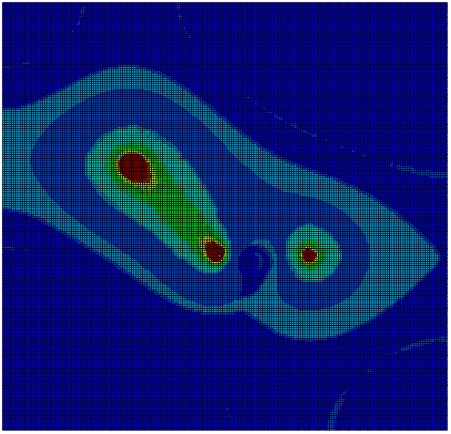} }
\subfigure[$t=20$]{ \includegraphics[scale=0.85]{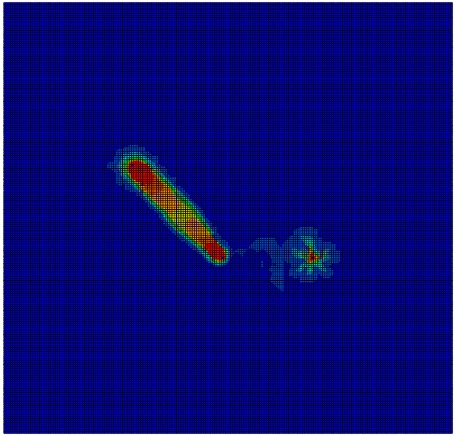} }
\subfigure[$t=30$]{ \includegraphics[scale=0.85]{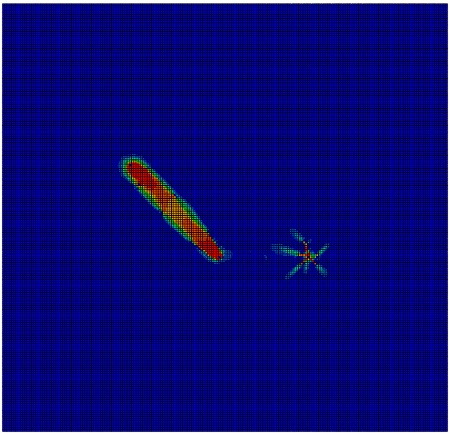} }
\subfigure[$t=40$]{ \includegraphics[scale=0.85]{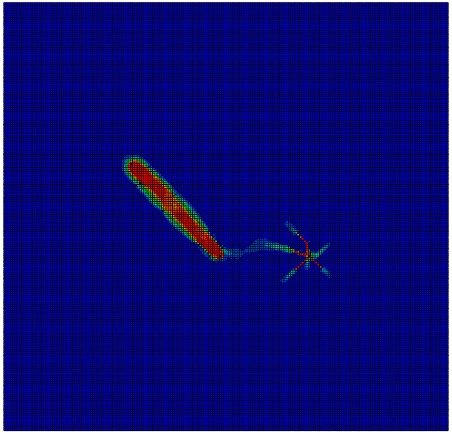} }
\subfigure[$t=50$]{ \includegraphics[scale=0.85]{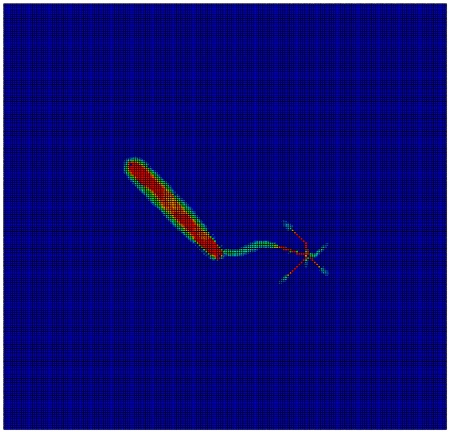} }
\subfigure[$t=200$]{ \includegraphics[scale=0.85]{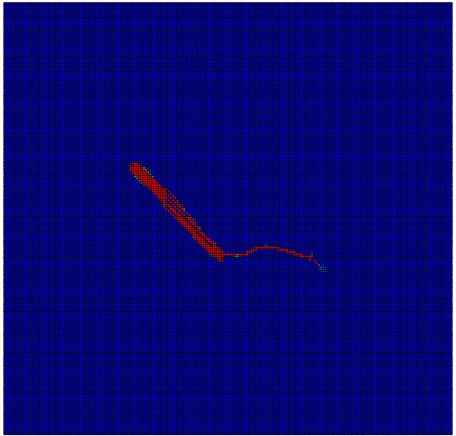} } 
\caption{Density distribution,  $\rho$, in the implementation of {\sc and} gate for inputs  $x=1$  and  $y=1$, 
Neumann boundary conditions for input points. The snapshots are taken at t=10, 20, 30, 40, 50, and 200 steps.}
\label{fig6}
\end{figure}

Let us consider the implementation of the {\sc and}  in case of Neumann boundary conditions for input points. 
Scheme of the gate is shown in Fig.~\ref{fig5}a.  The distance between $I_x$ and $I_y$ is 40 points, the distance between $I_x$ 
and $V$ is 70 points and between $I_y$ and outlet $V$ 90 points. The output site $O$ is positioned in the middle of the segment
$(I_x, I_y)$. Boundary conditions in $I_x$, $I_y$ and $V$ are set as fluxes, i.e. Neumann boundary conditions.

Figure~\ref{fig5}b shows density distribution, $\rho$ for inputs $x=1$ and $y=0$. The maximum density region
develops along the shortest path $(I_x, V)$. Therefore, the density value at $O$ is minimal,  $\rho_O=\rho_{\min}$, which represents
logical output {\sc False}.  For inputs $x=0$ and $y=1$ (Fig.\ref{fig5}c) the maximal density region is formed along the path 
$(I_y, V)$, i.e. $\rho_O=\rho_{\min}$ and the logical output is {\sc False}. The material density distribution for inputs $x=1$ and $y=1$ is shown in 
Fig.\ref{fig5}d.  The maximum density region develops along the path $(I_y, I_x, V)$. Thus $\rho_O=\rho_{\max}$ and logical output is {\sc True}.

Figure \ref{fig6} shows intermediate results of simulating density distribution, $\rho$, for inputs $x=1$ and $y=1$. At beginning of computation 
the material develops in proximity of $I_x$, $I_y$ and $V$ (Fig.~ \ref{fig6}a). Then $I_x$ and $V$ are connected by a domain with highest density of the material (Fig.~ \ref{fig6}b). The thinner region of high-density material is further develops between $I_x$ and $I_y$ (Fig.~ \ref{fig6}c--f).

\section{{\sc xor} gate}
\label{xorgate}

\subsection{Dirichlet boundary conditions}

\begin{figure}[!tbp]
\centering
\subfigure[]{\includegraphics[scale=0.24]{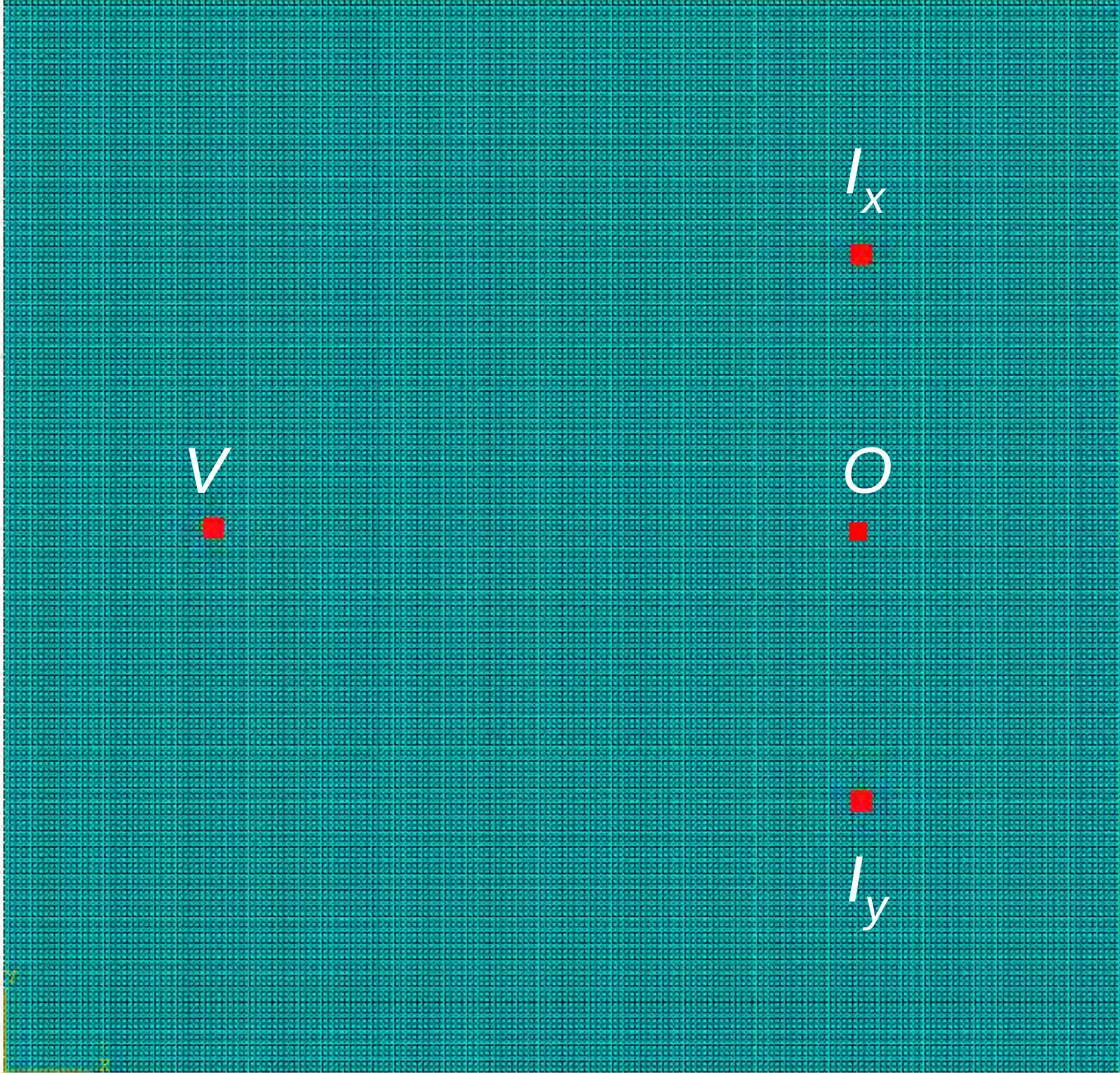} }
\subfigure[]{\includegraphics[scale=0.24]{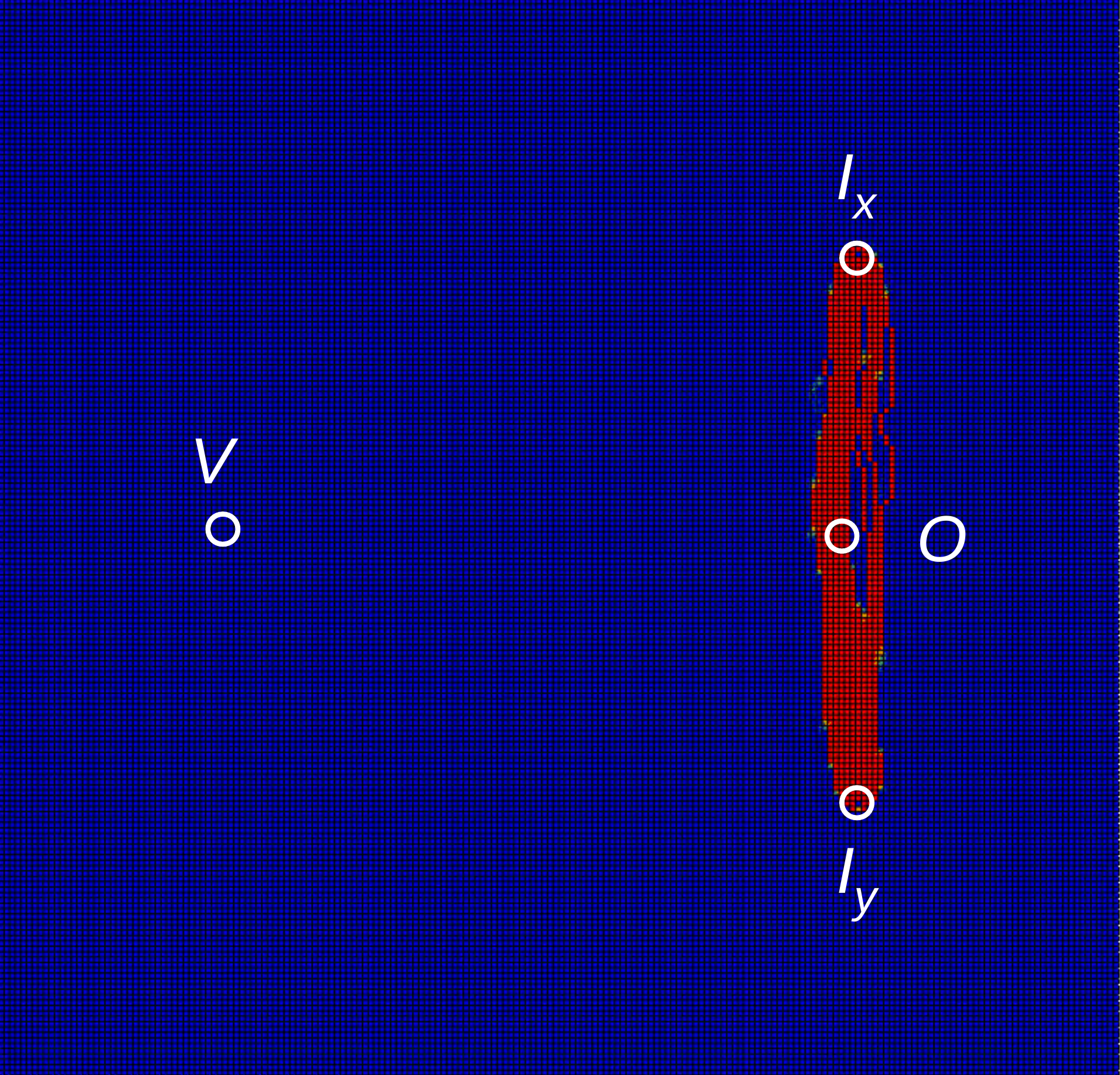} }
\subfigure[]{\includegraphics[scale=0.24]{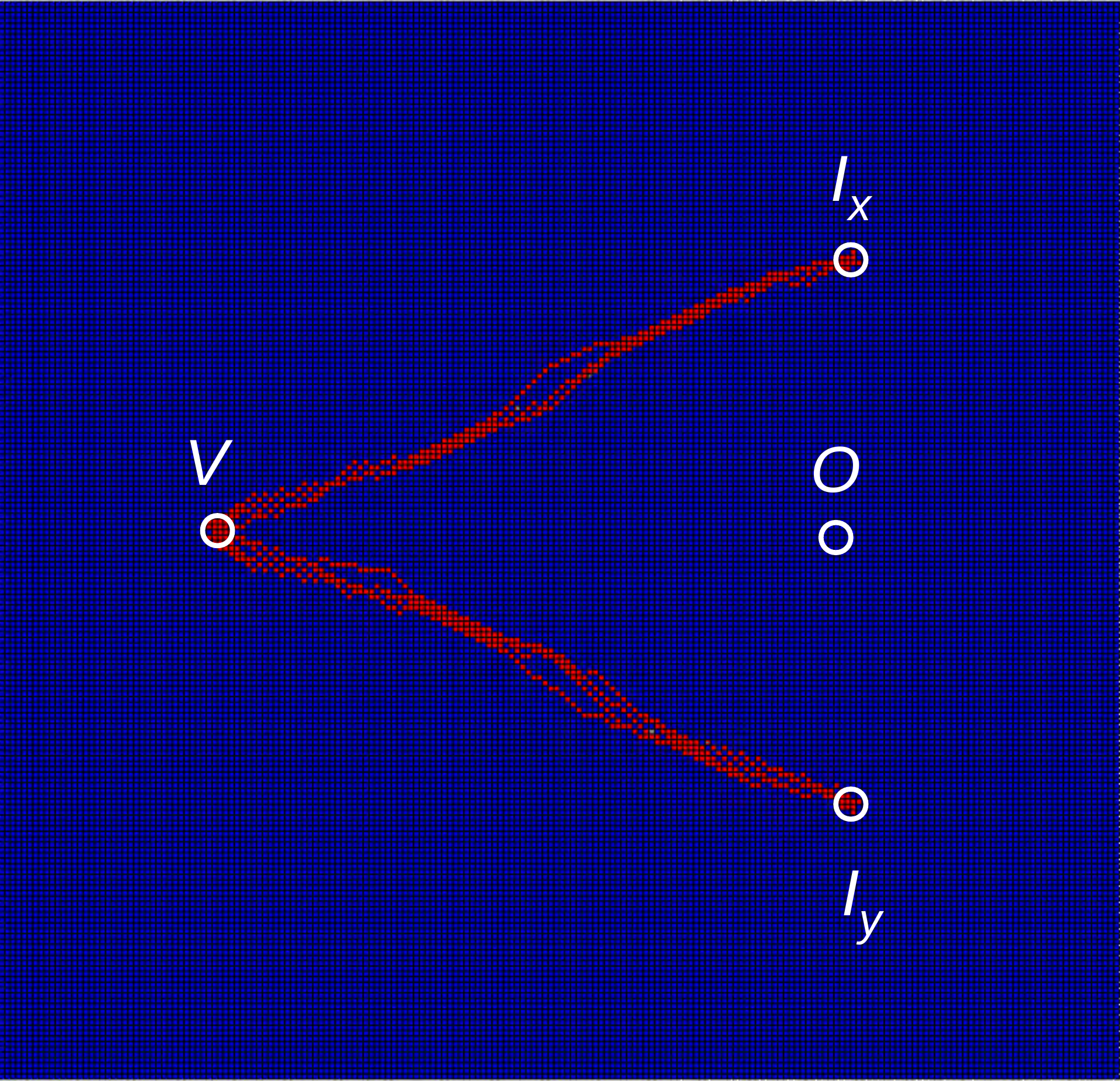} }
\caption{{\sc xor} gate implementation with Dirichlet boundary conditions. 
(a) Scheme of inputs and outputs. (bc) Density distribution $\rho$ for inputs (b) $x=1$ and $y=0$ and 
(c ) $x=1$ and $y=1$. }
\label{fig4}
\end{figure}

Let us consider the implementation of the {\sc xor} gate in case of Dirichlet boundary conditions for input points. We use similar design as in {\sc and} gate (Fig.~\ref{fig1}) but  use two inputs $I_x$ and $I_y$, output $O$  and outlet $V$. The site of output $O$ in {\sc and} gate is assigned outlet $V$ function
and the output site $O$ is positioned in the middle of the segment connecting sites $I_x$ and $I_y$ (Fig~\ref{fig4}a). 
The temperature at $V$ point is set to 0, $T_V=0$, no temperature boundary conditions are set at $O$. If only one input is {\sc True} a region of maximum density material is formed along a shortest path between $I_x$ and $I_y$. Therefore, the density value $\rho_O=\rho_{\max}$ thus indicated output {\sc True} (Fig.~\ref{fig4}b, $x=1$, $y=0$). When both inputs variables  are {\sc True}, $x=1$ and $y=1$, maximum density regions are formed along the path 
$(I_x, V)$ and $(I_y, V)$ not along $(I_x, I_y)$. Thus $\rho_0 = \rho_{\min}$, i.e. logical output {\sc False} (Fig.~\ref{fig4}c, $x=1$, $y=1$).

\subsection{Neumann boundary conditions.}

\begin{figure}[!tbp]
\centering
\subfigure[]{\includegraphics[scale=0.25]{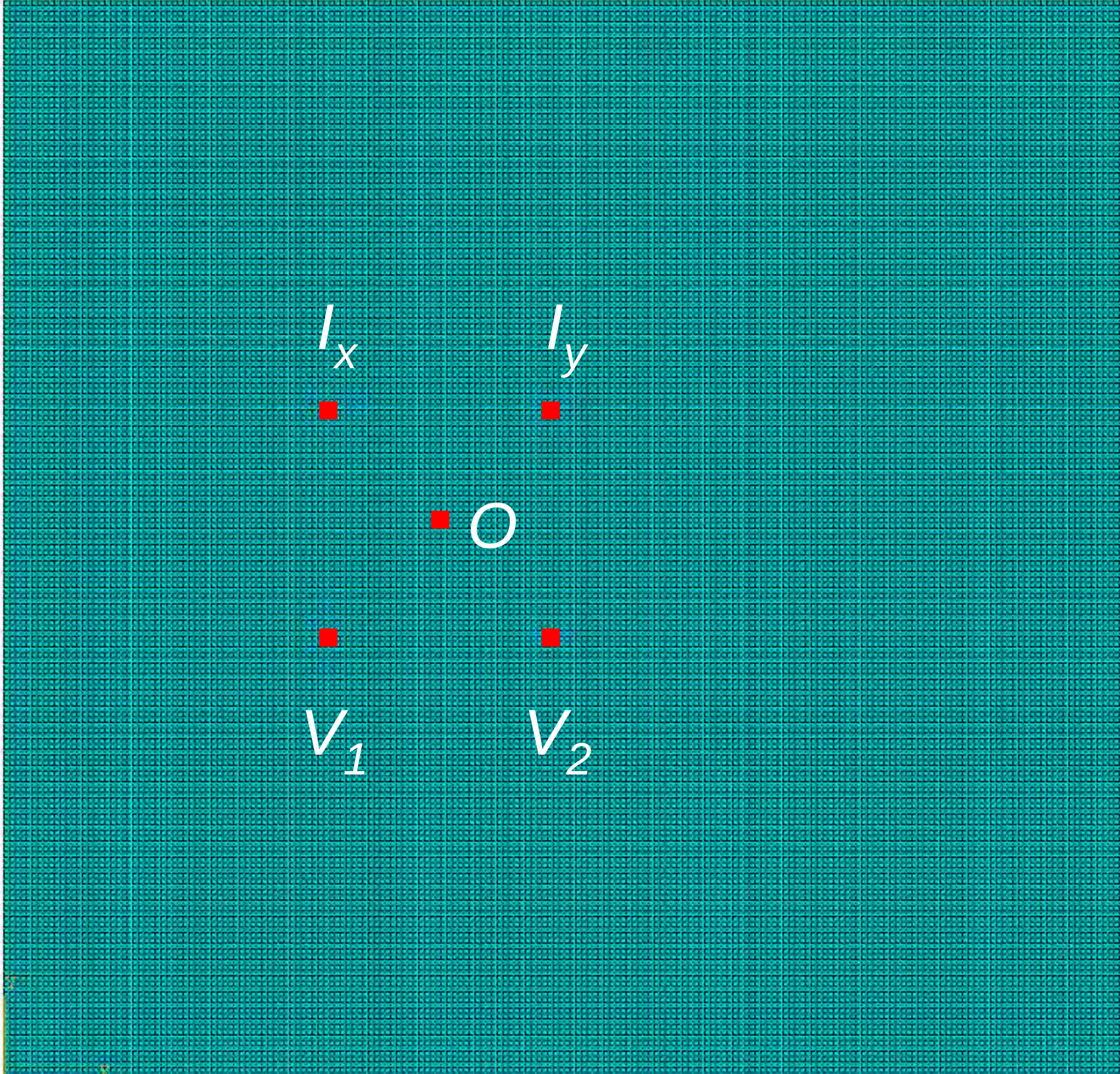}}
\subfigure[]{\includegraphics[scale=0.25]{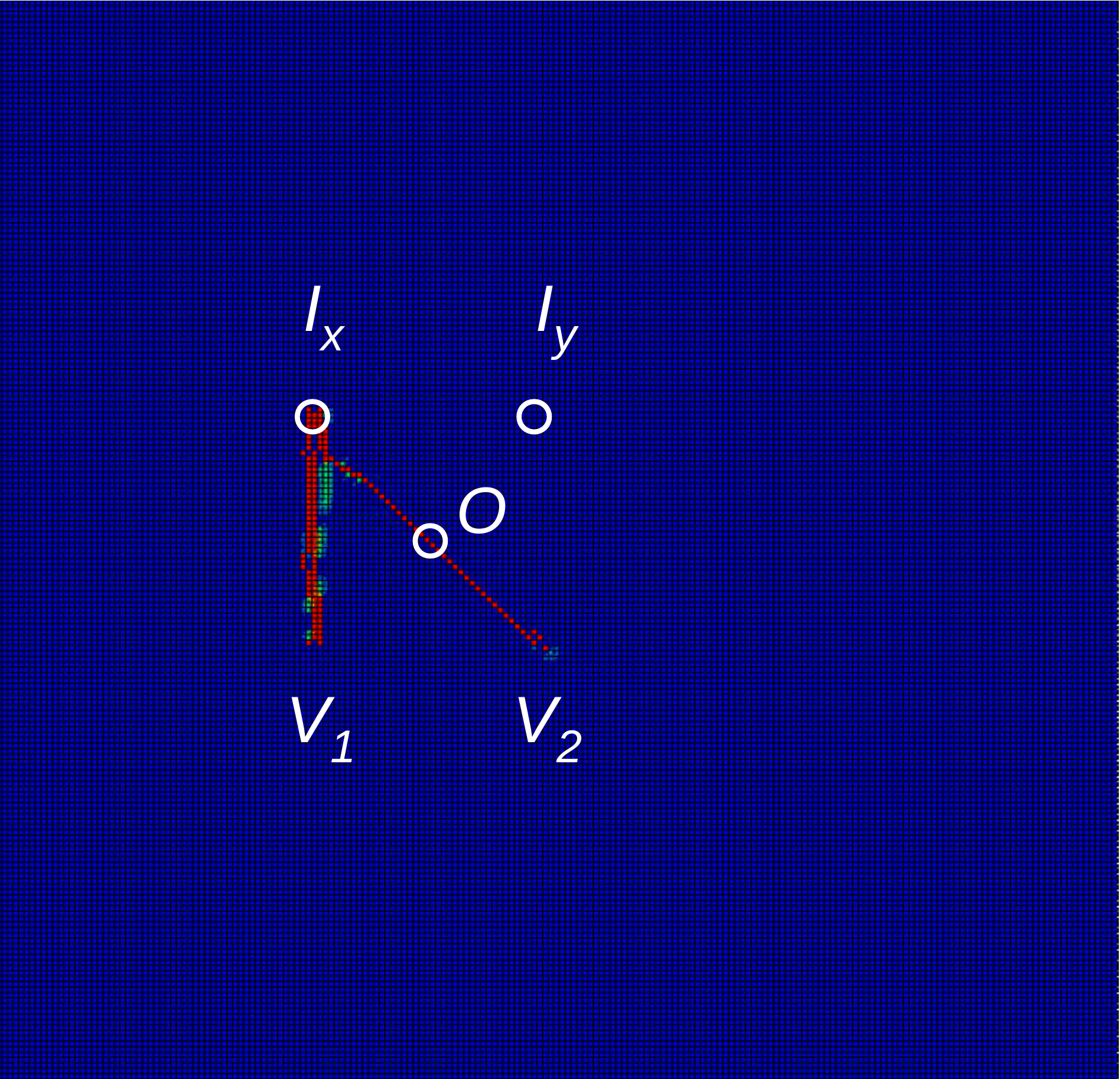}}\\
\subfigure[]{\includegraphics[scale=0.25]{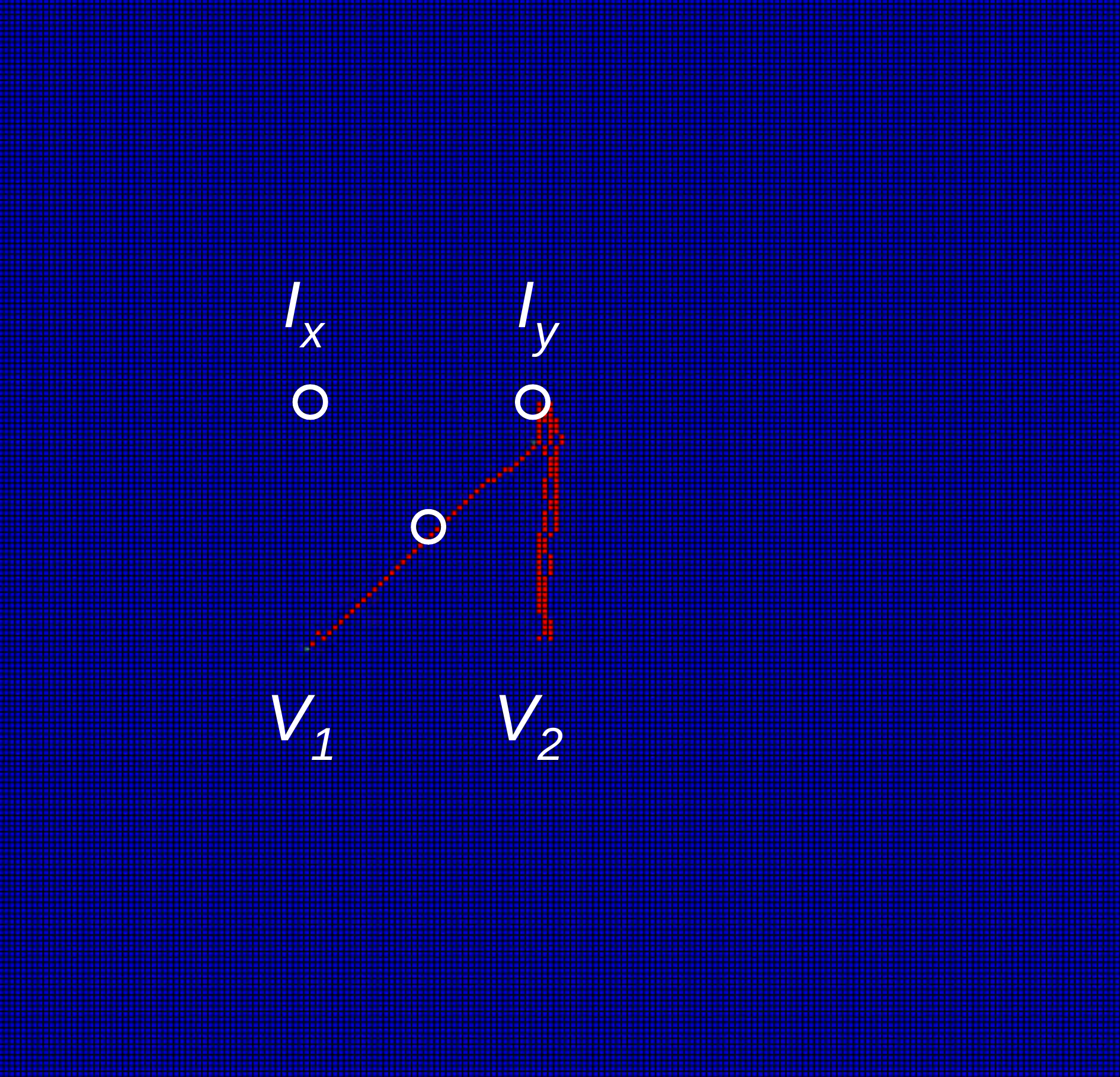}}
\subfigure[]{\includegraphics[scale=0.25]{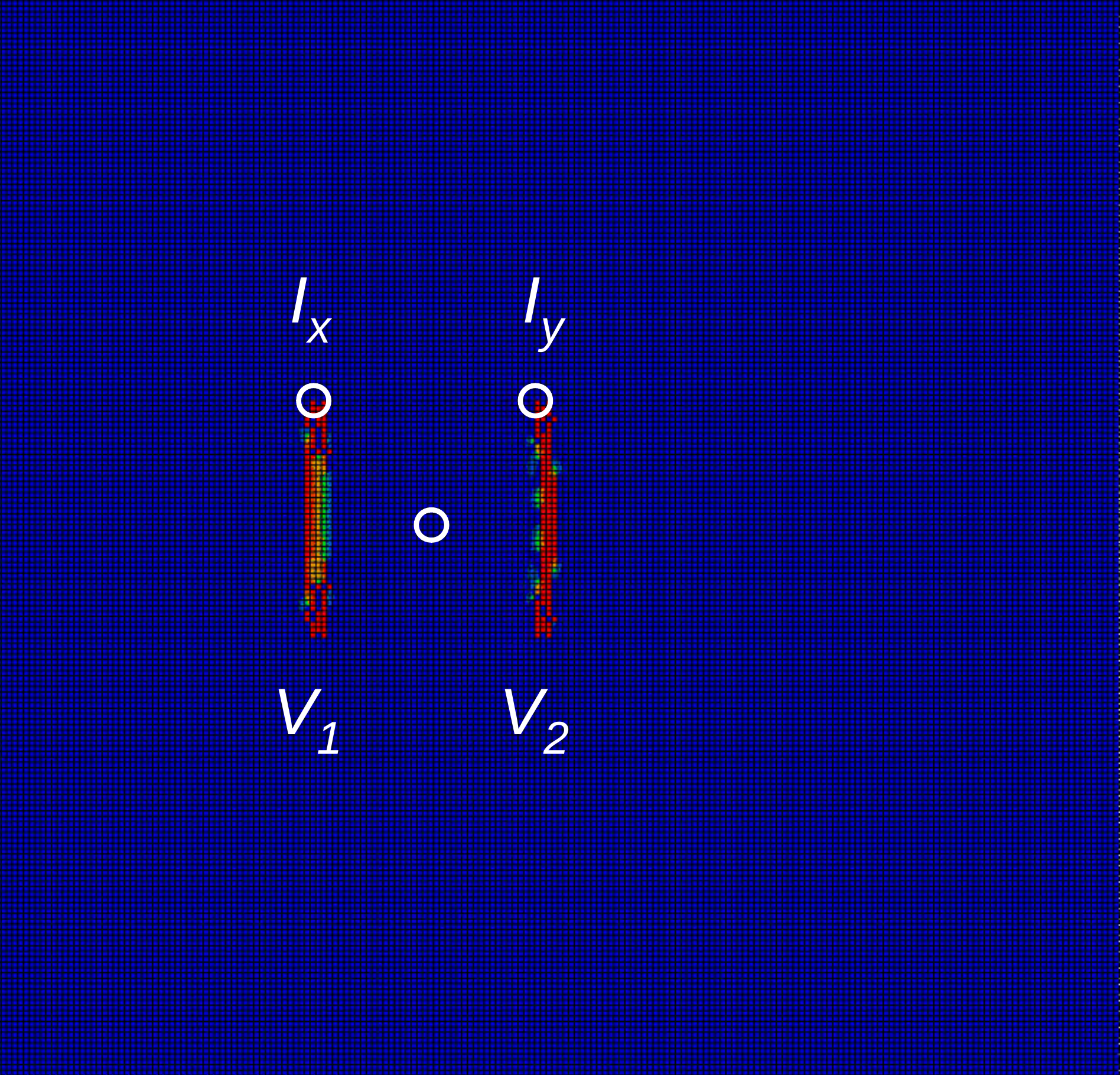}}
\caption{{\sc xor} gate implementation in case of Neumann boundary conditions.
(a) Scheme of the gate. Density distribution, $\rho$, for inputs
(b)  $x=1$, $y=0$,
(c )  $x=0$, $y=1$,
(d)  $x=1$, $y=1$.}
\label{fig7}
\end{figure}

Let us consider the implementation of {\sc xor} gate  in case of Neumann boundary conditions for input points. 
The gate has five sites: inputs $I_x$ and $I_y$, output $O$, outlets $V_1$ and $V_2$ (Fig.~\ref{fig7}a).  
Sites $I_x$, $I_y$, $V_1$ and $V_2$ are vertices of the square with the side length 42 points. The output site $O$
is positioned at the intersection of diagonals of the square. Boundary conditions in $I_x$, $I_y$, $V_1$ and $V_2$ 
are set as fluxes, i.e. Neumann boundary conditions. To ensure convergence of solutions for the stationary problem of 
heat conduction (1) the fluxes at $V_1$ and $V_2$  are set equal to the negative half-sum of fluxes in $I_x$ and $I_y$:
$Q_{V_1} = Q_{V_2} = - \frac{Q_{I_x}+Q_{I_y}}{2}$.

For $x=1$ and $y=0$ the maximum density domain is formed between $I_x$ and $V_1$  and 
between $I_x$ and $V_2$ (Fig~\ref{fig7}b). The output site $O$ sits at the $(I_x, V_2)$ diagonal, 
therefore $\rho_O = \rho_{\max}$, and thus the logical  output is {\sc True}. When inputs are $x=0$ and $y=1$
the maximum density domain is formed between $I_y$ and $V_2$  and  between $I_y$ and $V_1$ (Fig~\ref{fig7}c).
The output site $O$ sits at the $(I_y, V_1)$ diagonal, therefore $\rho_O = \rho_{\max}$, and thus the logical output is {\sc True}.
When goths inputs are {\sc True}, $x=1$ and $y=1$, domains of high-density material develop along shortest paths 
$(I_x, V_1)$ and $(I_y, V_2)$ (Fig~\ref{fig7}d). These domain do not cover the site $O$, therefore logical output is {\sc False}.

\section{One-bit half-adder}
\label{onebithalfadder}

\subsection{Dirichlet boundary conditions}

\begin{figure}[!tbp]
\centering
\subfigure[]{\includegraphics[scale=0.25]{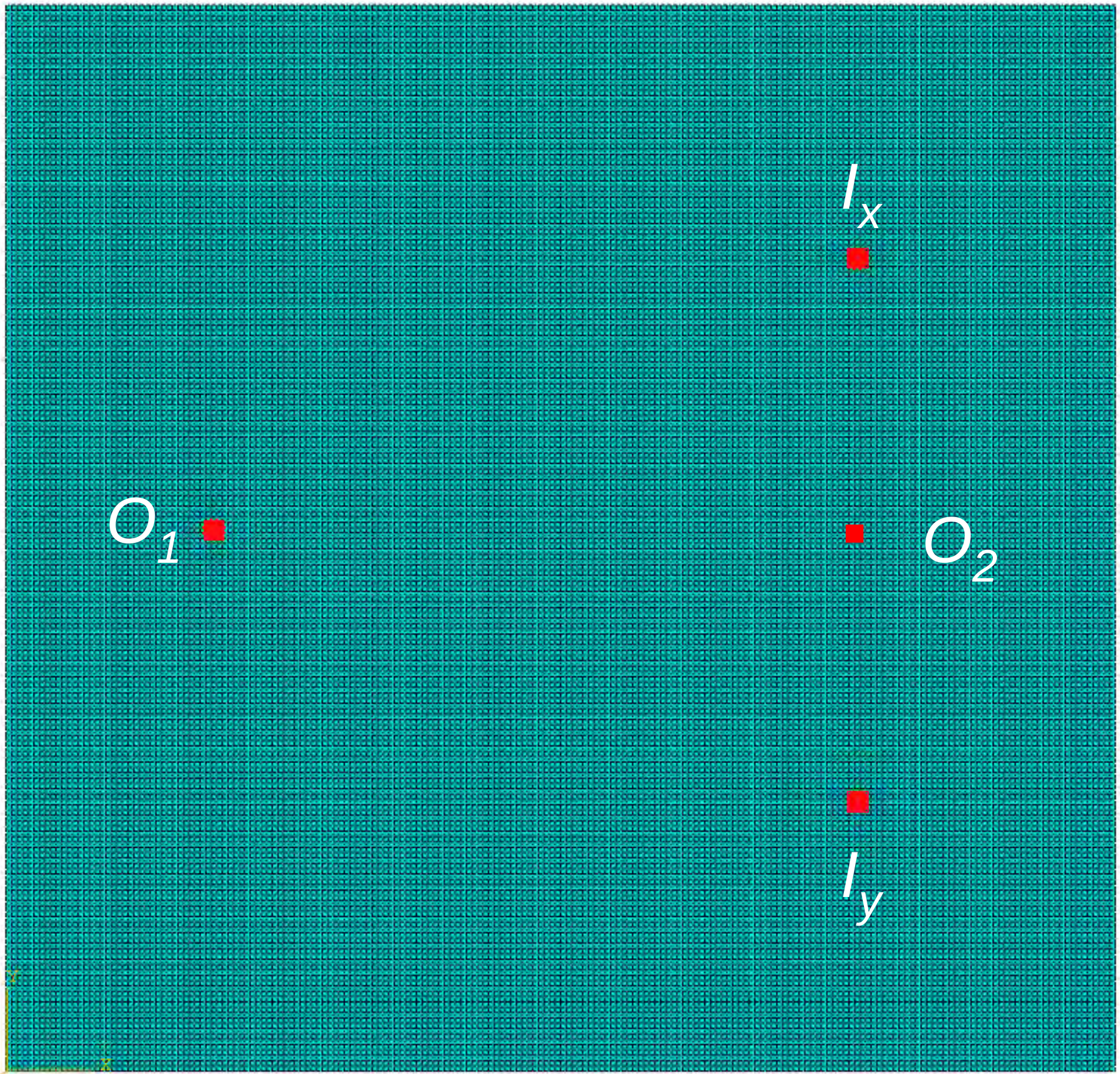}}
\subfigure[]{\includegraphics[scale=0.25]{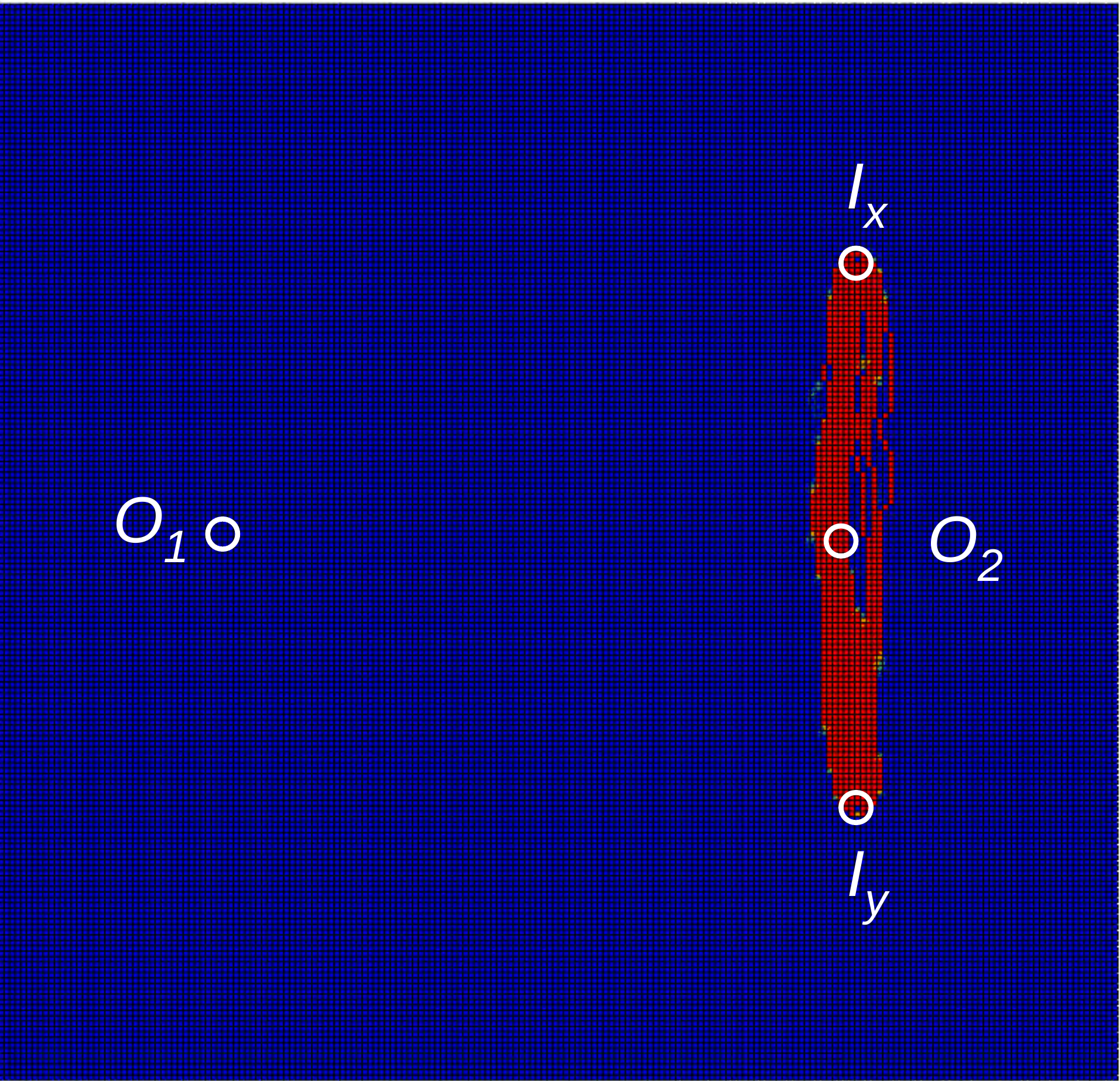}}
\subfigure[]{\includegraphics[scale=0.25]{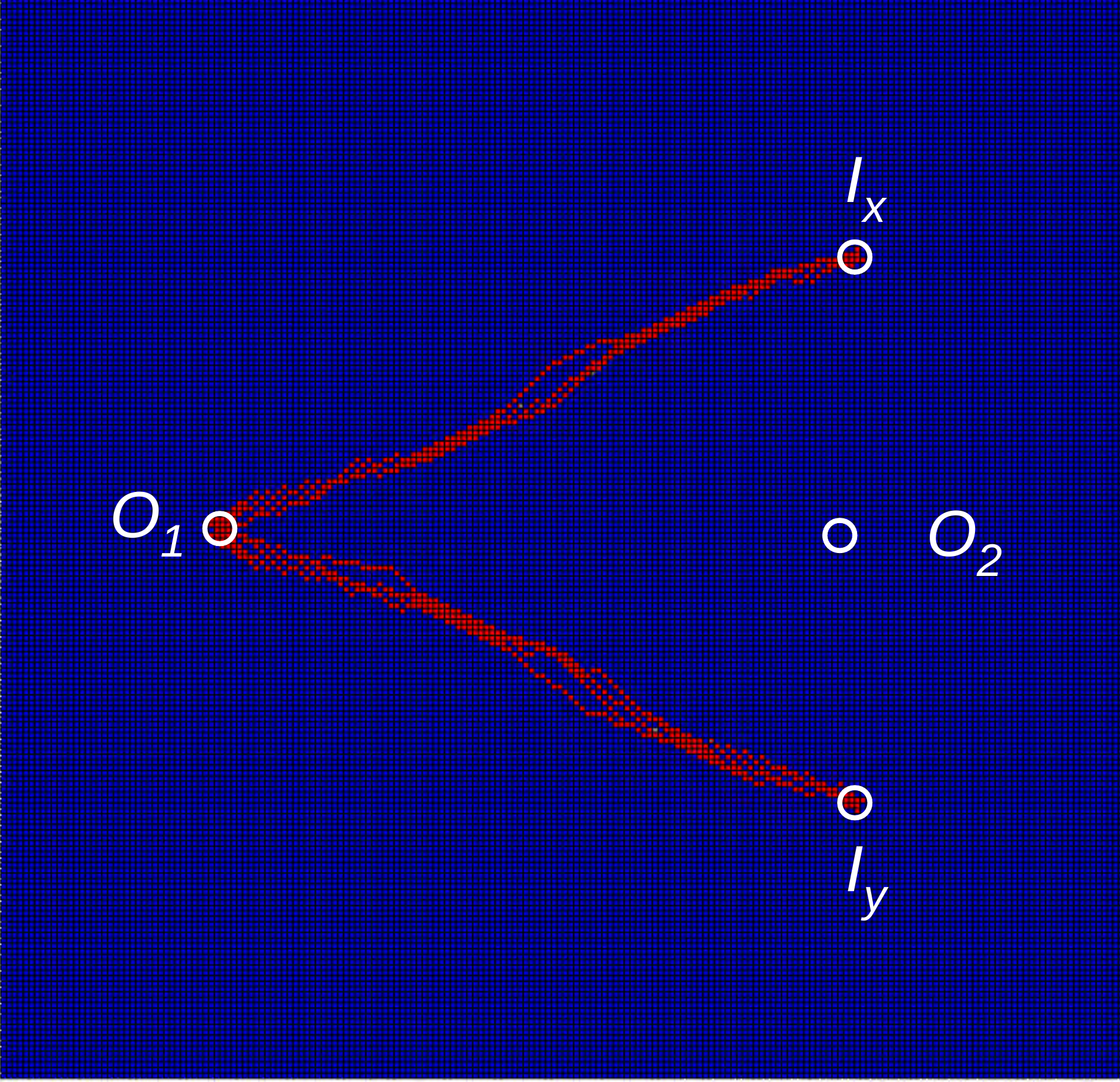}}
\caption{One-bit half-adder implementation in case of Dirichlet boundary conditions.
(a) Scheme of the adder. Density distribution, $\rho$, for inputs
(b)  $x=1$, $y=0$,
(c )  $x=1$, $y=1$.
}
\label{fig8}
\end{figure}

To implement the one-bit  half-adder in case of Dirichlet boundary conditions for input points we combine designs of {\sc and} and {\sc xor} gates
(Figs.~\ref{fig1}a and \ref{fig4}a). We introduce the following changes to the scheme shown in Fig~\ref{fig4}a:  the former outlet $V$
is designated as output $O_1$, the former output $O$ is designated as output $O_2$ (Fig.~\ref{fig8}a).  Temperature value at $O_1$ is set zero, 
$T_{O_1}=0$. No temperature boundary conditions are set at $O_2$. The output $O_1$ indicated logical value $xy$ and the output $O_2$ logical 
value $x \oplus y$. When only one of the inputs is {\sc True} and other {\sc False}, e.g. $x=1$ and $y=0$ as shown in Fig.~\ref{fig8}b, the density
value at $O_1$ is minimal, $\rho_{O_1}=\rho_{\min}$, and the density value at $O_2$ is maximal, $\rho_{O_2}=\rho_{\max}$.  Thus $O_1$ indicated {\sc False} and $O_2$ {\sc True}. For inputs $x=1$ and $y=1$ we have $\rho_{O_1}=\rho_{\max}$ and $\rho_{O_2}=\rho_{\min}$, i.e. logical outputs 
{\sc True} and {\sc False}, respectively.

\subsection{Neumann boundary conditions}

\begin{figure}[!tbp]
\centering
\subfigure[]{\includegraphics[scale=0.25]{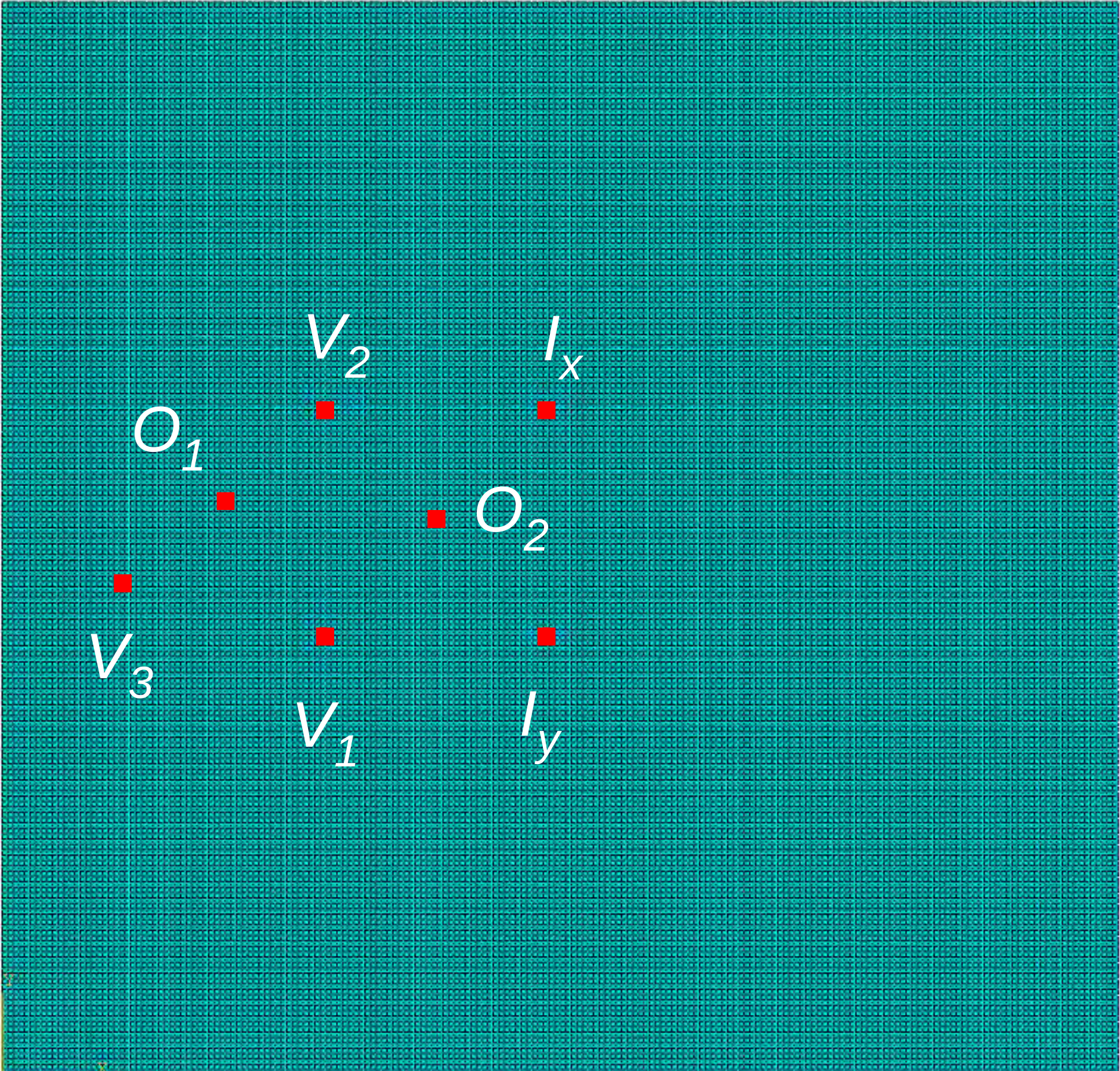}}
\subfigure[]{\includegraphics[scale=0.25]{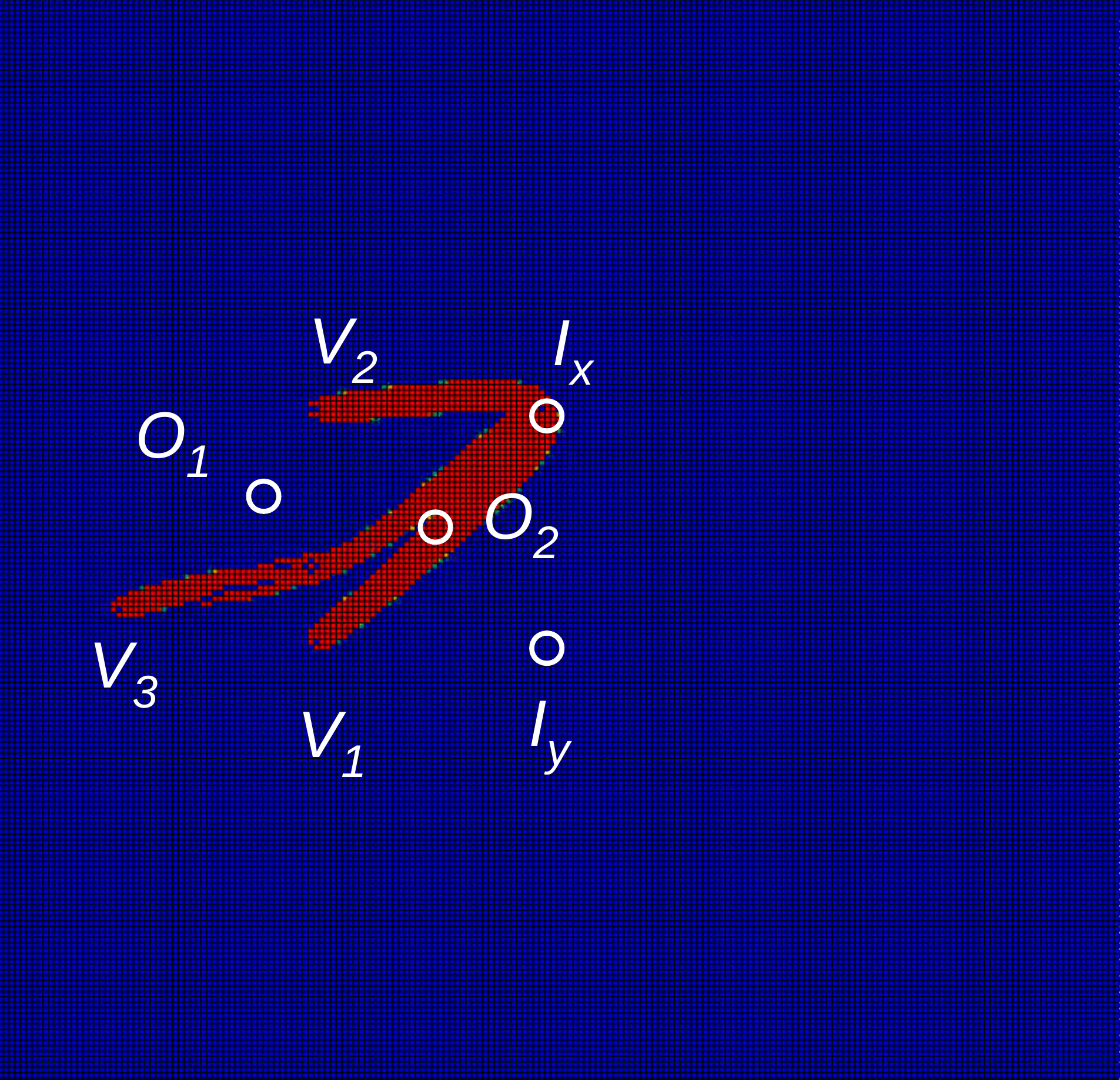}}\\
\subfigure[]{\includegraphics[scale=0.25]{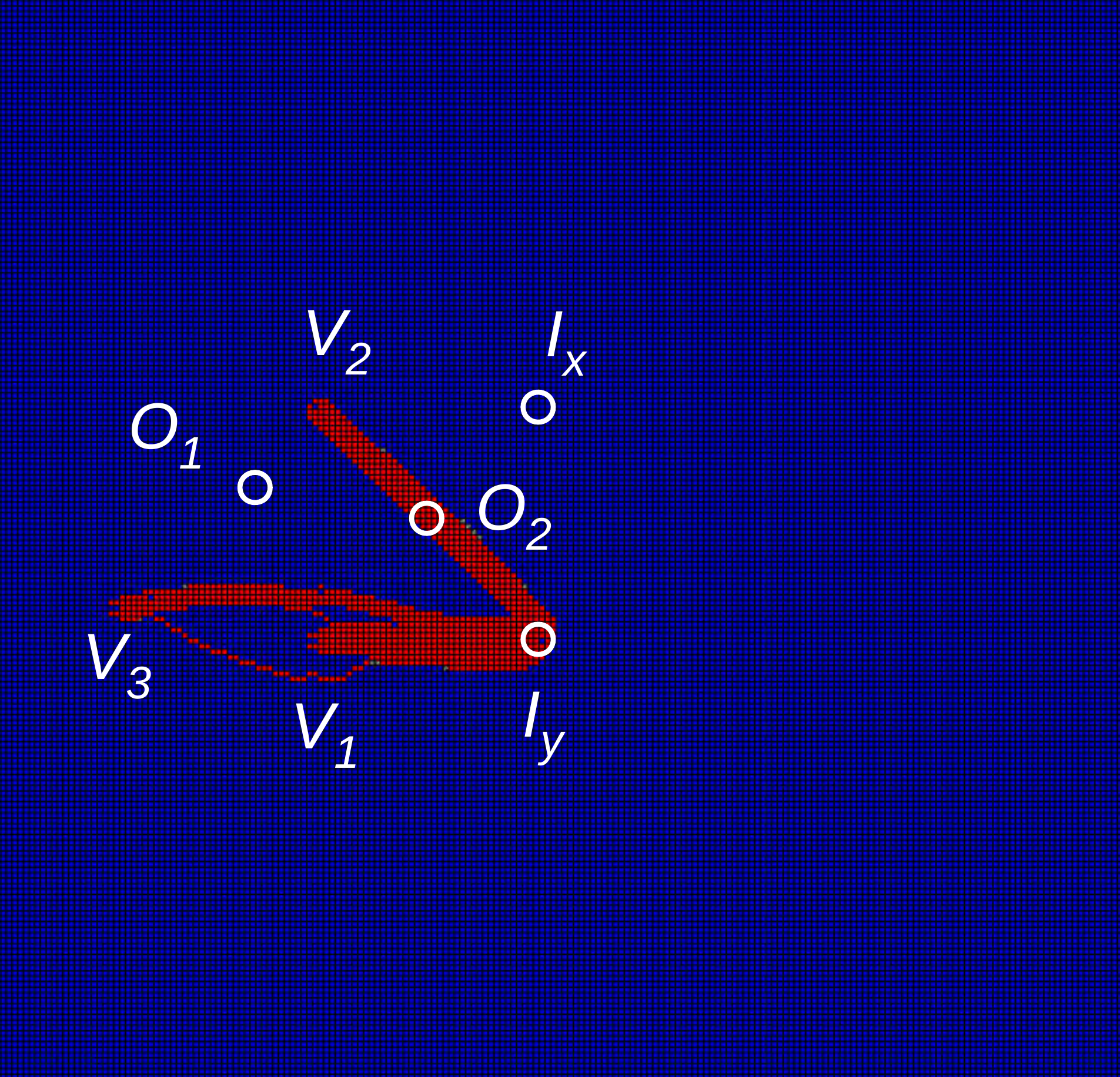}}
\subfigure[]{\includegraphics[scale=0.25]{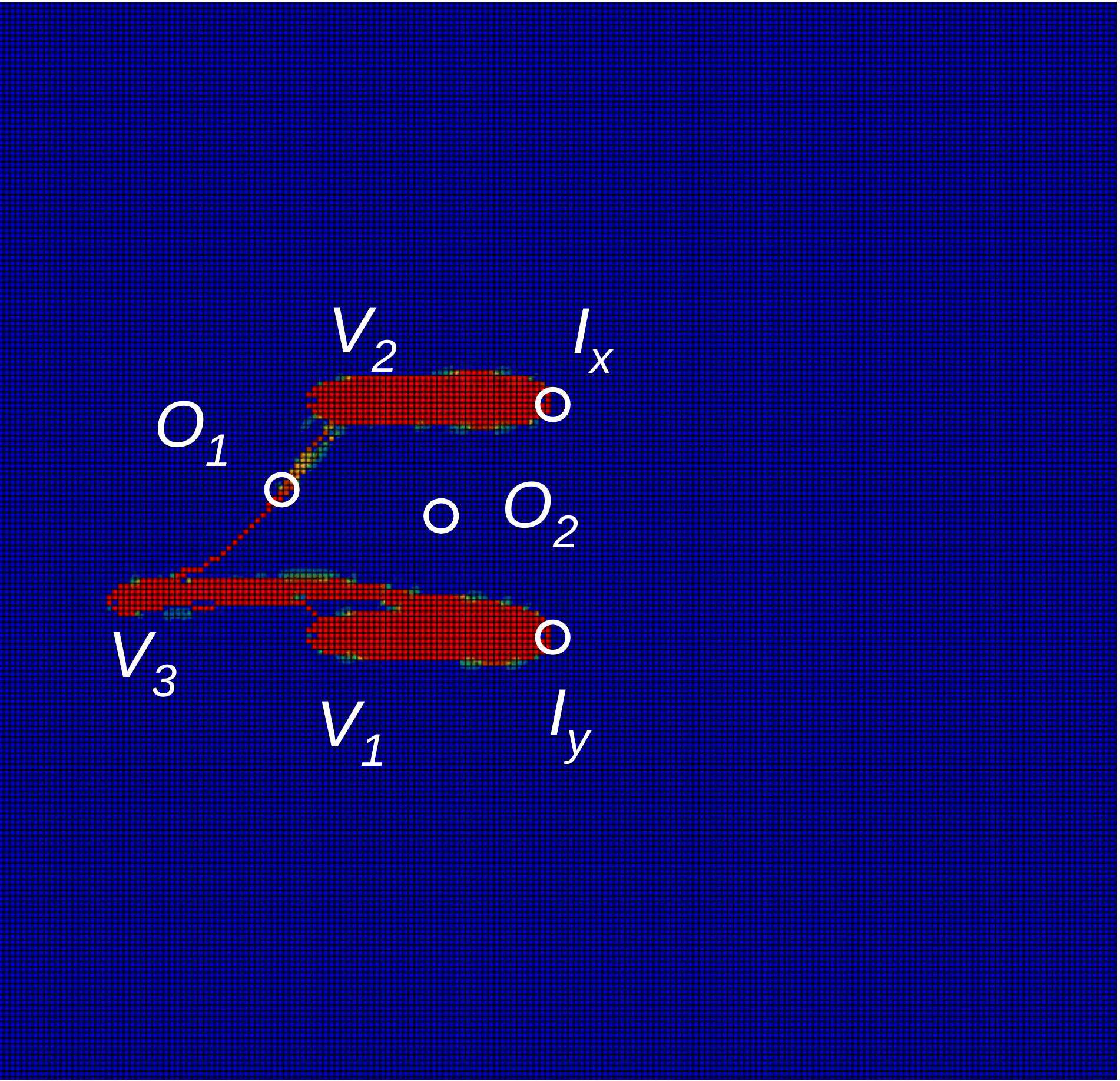}}
\caption{One-bit half-adder implementation in case of Neumann boundary conditions.
(a) Scheme of the adder. Density distribution, $\rho$, for inputs
(b)  $x=1$, $y=0$,
(c ) $x=0$, $y=1$,
(b)  $x=1$, $y=1$.
}
\label{fig9}
\end{figure}

\begin{figure}[!tbp]
\centering
\subfigure[$t=10$]{ \includegraphics[scale=0.85]{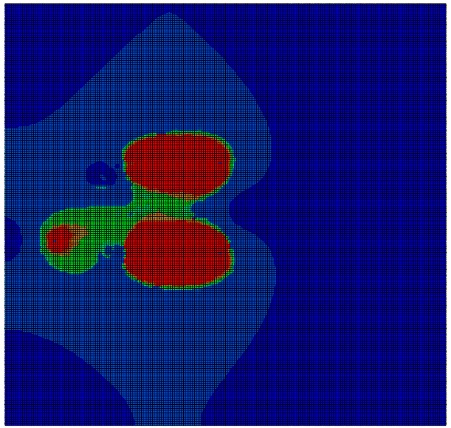} }
\subfigure[$t=20$]{ \includegraphics[scale=0.85]{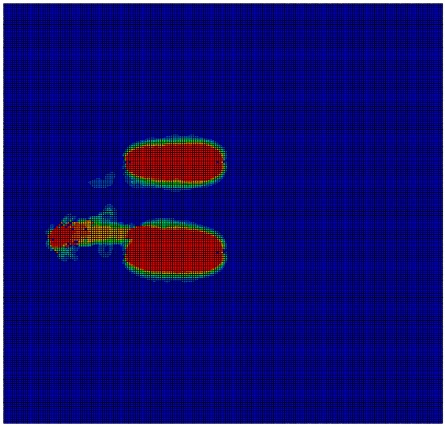} }
\subfigure[$t=30$]{ \includegraphics[scale=0.85]{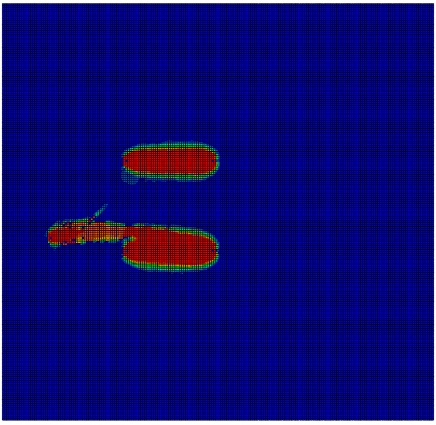} }
\subfigure[$t=40$]{ \includegraphics[scale=0.85]{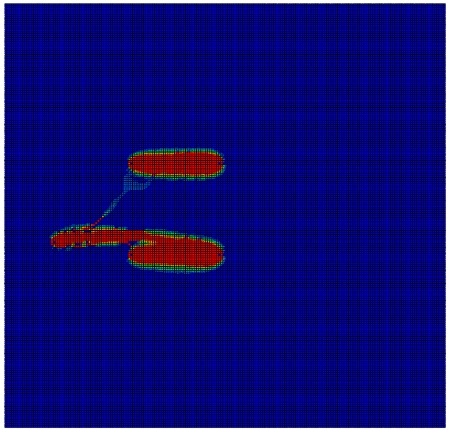} }
\subfigure[$t=50$]{ \includegraphics[scale=0.85]{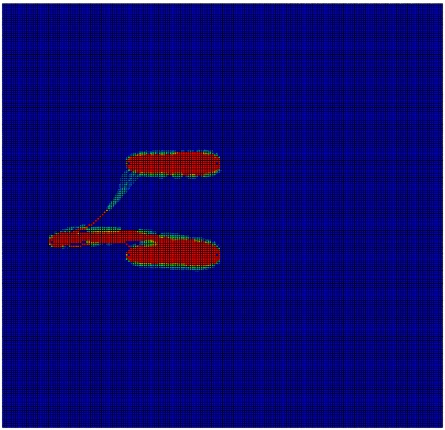} }
\subfigure[$t=69$]{ \includegraphics[scale=0.85]{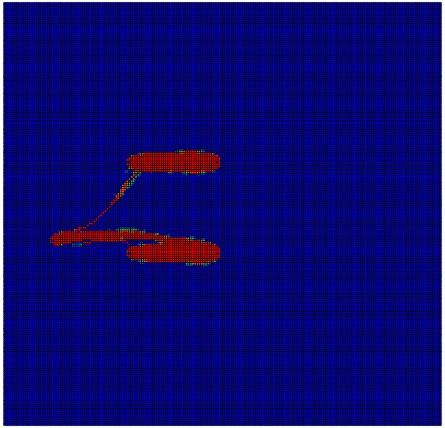} } 
\caption{Density distribution,  $\rho$, in the implementation of one-bit half-adder for inputs  $x=1$  and  $y=1$, 
Neumann boundary conditions for input points. The snapshots are taken at t=10, 20, 30, 40, 50, and 69 steps.}
\label{fig10}
\end{figure}

Let us consider the implementation of one-bit half-adder in case of Neumann boundary conditions for input points. 
The devices consists of seven sites: two inputs $I_x$ and $I_y$, two outputs $O_1$ and $O_2$, 
three outlets $V_1$, $V_2$ and $V_3$ (Fig.~\ref{fig9}a). Sites $I_x$, $I_y$, $O_1$ and $O_2$ are vertices of a square with 
the side length 40. The output $O_2$ is positioned at the intersection of diagonals of this square. The output $O_1$ is positioned 
at the middle of the segment connecting $V_2$ and $V_3$. The distance between $V_1$ and $V_3$ is 36 points, the distance between 
$V_3$ and $V_2$ is 51 points. The output $O_1$ represent logical function $xy$ and the output $O_2$ function $x \oplus y$. 
Boundary conditions in $I_1$, $I_2$, $V_1$, $V_2$ and $V_3$ are set as fluxes,  thus corresponding to Neumann boundary 
conditions. To ensure convergence of solutions for the stationary problem of heat conduction (1) the flux values at
$V_1$, $V_2$ and $V_3$ are set equal to one third of the negative sum of fluxes in  $I_x$ and $I_2$: 
$Q_{V_1} = Q_{V_2} = Q_{V_3} = - \frac{Q_{I_x}+Q_{I_y}}{3}$.

Figure~\ref{fig9}b shows results of calculating density distribution $\rho$ for inputs $x=1$ and $y=0$. There the 
maximum density region connects $I_x$ with $V_1$, $V_2$ and $V_3$. The density domain $(I_x, V_3)$ is not a straight line because
the system benefits most when a of the segment $(I_x, V_3)$ coincide with the segment $(I_x, V_1)$. The site $O_2$ is covered by maximum 
density domain $(I_x, V_1)$, $\rho_{O_2} = \rho_{\max}$, thus representing logical value {\sc True}; the output $O_1$ is {\sc False} because
$\rho_{O_2} = \rho_{\min}$. 

The density distribution $\rho$ calculated for inputs $x=0$ and $y=1$ is shown in Fig.~\ref{fig9}c. The maximum density region 
connects $I_y$ with $V_1$, $V_2$ and $V_3$ via paths $(I_y, V_1)$, $(I_y, V_2)$, $(I_y, V_3)$. The output $O_2$ belongs to $(I_y, V_2)$ therefore it 
indicates logical output {\sc True}. The output $O_1$ indicated {\sc False} because it is not covered by a high density domain. 

The density distribution $\rho$ calculated for inputs $x=1$ and $y=1$ is shown in Fig.~\ref{fig9}d. The maximum density regions are developed along paths
$(I_1, V_2)$, $(I_2, V_1)$, $(I_2, V_3)$ and $(V_2, V_3)$. There is heat flux between $V_1$ and $V_2$ which forms a segment of high density material.
The high density material covers $O_1$, therefore the output $O_1$ indicate logical value {\sc True}. The output $O_2$ is not covered by a high density material, 
thus {\sc False}.

Figure \ref{fig10} shows intermediate results of simulating density distribution $\rho$ for inputs $x=1$ and $y=1$.

\section{Discussion}
\label{discussion}

We implemented logical gates and circuits using optimisation of conductive material when solving stationary problems of heat conduction. 
In the simplest case of two sites with given heat fluxes the conductive material is distributed between the sites in a straight line. 
The implementations of gates presented employ several sites, exact configuration of topologically optimal structures of the conductive 
material is determined by value of input variables. The algorithm of optimal layout of the conductive material is similar to a biological process of 
bone remodelling. The algorithm proposed can be applied to a wide range of biological networks, including  neural networks, 
vascular networks, slime mould, plant routes, fungi mycelium. These networks will be the subject of further studies. In future we can also 
consider an experimental laboratory testing of the numerical implementations of logical gates, e.g. via dielectric breakdown tests, because
the phenomenon is also described by Laplace's stationary heat conduction equation which takes into account the evolution of conductivity of 
a medium determined by the electric current. The approach to developing logical circuits, proposed by us, could be used in fast-prototyping of 
experimental laboratory unconventional computing devices. Such devices will do computation by changing properties of their material substrates. 
First steps in this direction have been in designing Belousov-Zhabotinsky medium based computing devices for pattern recognition~\cite{fang2016pattern} and 
configurable logical gates~\cite{wang2016configurable}, learning slime mould chip~\cite{whiting2016towards}, 
electric current based computing~\cite{ayrinhac2014electric},
programmable excitation wave propagation in living bioengineered tissues~\cite{mcnamara2016optically}, heterotic computing~\cite{kendon2015heterotic}, memory devices in digital collides~\cite{phillips2014digital}.

\section*{Supplemetary materials}

\subsection*{{\sc xor} gate, Neumann boundary conditions}
\begin{itemize}
\item inputs $x=0$, $y=1$: \url{https://www.youtube.com/watch?v=osB12UqM3-w}
\item inputs $x=1$, $y=0$: \url{https://www.youtube.com/watch?v=lKMeu1nFuak}
\item inputs $x=1$, $y=1$: \url{https://www.youtube.com/watch?v=AxdCVVtIqgk}
\end{itemize}

\subsection*{One-bit half-adder, Neumann boundary conditions} 
\begin{itemize}
\item inputs $x=0$, $y=1$: \url{https://www.youtube.com/watch?v=i81WTCrg8Lg}
\item inputs $x=1$, $y=0$: \url{https://www.youtube.com/watch?v=impbwJXjCAM}
\item inputs $x=1$, $y=1$: \url{https://www.youtube.com/watch?v=ubrgfzlAQQE}
\end{itemize}

\bibliographystyle{elsarticle-num}
\bibliography{bibliography}

\end{document}